\documentclass[english,prb,twocolumn,superscriptaddress]{revtex4-1}
\usepackage[T1]{fontenc}
\usepackage[latin9]{inputenc}
\usepackage{textcomp}
\usepackage{amsmath}
\usepackage{amssymb}
\usepackage{graphicx}
\usepackage{babel}
\begin{document}

\title{Dynamics of the superconducting order parameter through ultrafast
normal-to-superconducting phase transition in Bi$_{2}$Sr$_{2}$CaCu$_{2}$O$_{8+\delta}$
from multi-pulse polarization-resolved transient optical reflectivity}

\author{I. Madan}

\affiliation{Complex Matter Department, Jozef Stefan Institute, Jamova 39, 1000
Ljubljana, Slovenia}

\affiliation{Laboratory for Ultrafast Microscopy and Electron Scattering, IPHYS,
École Polytechnique Fédérale de Lausanne, CH-1015 Lausanne, Switzerland}

\author{V.V. Baranov}

\affiliation{Complex Matter Department, Jozef Stefan Institute, Jamova 39, 1000
Ljubljana, Slovenia}

\author{Y. Toda}

\affiliation{Department of Applied Physics, Hokkaido University, Sapporo 060-8628,
Japan}

\author{M. Oda}

\affiliation{Department of Physics, Hokkaido University, Sapporo 060-0810, Japan}

\author{T. Kurosawa}

\affiliation{Department of Physics, Hokkaido University, Sapporo 060-0810, Japan}

\author{V.V. Kabanov}

\affiliation{Complex Matter Department, Jozef Stefan Institute, Jamova 39, 1000
Ljubljana, Slovenia}

\author{T. Mertelj}

\email{tomaz.mertelj@ijs.si}

\affiliation{Complex Matter Department, Jozef Stefan Institute, Jamova 39, 1000
Ljubljana, Slovenia}

\affiliation{Center of Excellence on Nanoscience and Nanotechnology Nanocenter
(CENN Nanocenter), Jamova 39, 1000 Ljubljana, Slovenia}

\author{D. Mihailovic}

\affiliation{Complex Matter Department, Jozef Stefan Institute, Jamova 39, 1000
Ljubljana, Slovenia}

\affiliation{Center of Excellence on Nanoscience and Nanotechnology Nanocenter
(CENN Nanocenter), Jamova 39, 1000 Ljubljana, Slovenia}
\begin{abstract}
A systematic temperature dependent study of the femtosecond optical
superconducting (SC) state destruction and recovery in Bi$_{2}$Sr$_{2}$CaCu$_{2}$O$_{8+\delta}$
cuprate superconductor by means of the all-optical polarization-sensitive
multi-pulse spectroscopy is presented. At low temperatures and a partial
SC state suppression an anisotropic SC-gap recovery-timescale is suggested
by the data. The SC state destruction and recovery dynamics are compared
to the recent TR-ARPES-inferred SC-gap dynamics and a qualitative
agreement is found. Using a phenomenological response function the
experimental data are also compared to time dependent Ginzburg-Landau
model simulations. 
\end{abstract}
\maketitle

\section{Introduction}

The study of the time evolution of complex systems through symmetry
breaking transitions (SBT) is of great fundamental interest in different
areas of physics\cite{Bunkov2000,Higgs1966,Volovik2003}. An SBT of
particular general interest is the ultrafast normal-to-superconducting
(N$\rightarrow$S) state transition. Due to the small heat capacity
of the electronic system, an optical pulse can efficiently suppress
the SC state without heating the low-frequency phonon heat bath, which
remains well below the critical temperature ($T_{\mathrm{c}})$. This
enables us to perform an ultrafast effective\footnote{The quasiparticle energy distribution is nonthermal so strictly speaking
the electronic $T$ is not well defined.} electron temperature quench across $T_{\mathrm{c}}$ with an ultrashort
laser pulse, which is then followed by an ultrafast non-equilibrium
N$\rightarrow$S transition.

The ultrafast S$\rightarrow$N$\rightarrow$S transition in the cuprate
superconductors has been initially studied by all-optical\cite{CarnahanKaindl2004,Kusar2008,giannetti2009discontinuity,toda2011quasiparticle,BeyerStaedter2011}
pump-probe technique followed by laser ARPES\cite{SmallwoodHinton2012,SmallwoodZhang2014,SamallwoodZhang2015,PioveraZhang2015}.
While the laser ARPES can directly resolve the momentum dependent\footnote{Limited to the vicinity to the nodal point on the $\Gamma$-Y line.}
quasiparticle (QP) distribution function, all-optical techniques offer
better bulk sensitivity and greater flexibility. The lack of momentum
resolution of an optical probe can be partially compensated by use
of the optical dipole transition selection rules that depend on the
probe-photon polarization\cite{DevereauxHackl07,Toda2014} and energy\cite{toda2011quasiparticle,CoslovichGianetti2013}
and enable selection of different parts of the Brillouin zone (BZ). 

The electronic Raman-scattering tensor analyses have shown\cite{DevereauxHackl07}
that the dielectric tensor fluctuations of different symmetries can
be linked to charge excitations in different parts of the BZ. In particular,
in a D$_{\mathrm{4h}}$ point-symmetry corresponding to the ideal
CuO$_{2}$-plane symmetry, the dielectric tensor fluctuations with
the B$_{1\mathrm{g}}$ and B$_{2\mathrm{g}}$ symmetries are linked
to the anti-nodal and nodal BZ charge excitations, respectively, while
the totally symmetric A$_{1\mathrm{g}}$ fluctuations do not discriminate
between the regions. The transient reflectivity, $\Delta R$, is related
to the Raman tensor and in Bi$_{2}$Sr$_{2}$CaCu$_{2}$O$_{8+\delta}$
(Bi2212) the B$_{\mathrm{1g}}$-like\footnote{Despite Bi2212 is orthorhombic we use the ideal D$_{\mathrm{4h}}$
point group tetragonal CuO$_{2}$-plane symmetry to simplify the notation.
See supplemental to Ref. {[}\onlinecite{Toda2014}{]} for details.} transient reflectivity component shows sensitivity to the SC state
only, while A$_{1\mathrm{g}}$-like and B$_{2\mathrm{g}}$-like transient
reflectivity components couple to both the SC and pseudogap (PG) order.\cite{Toda2014}
\begin{figure}[t]
\includegraphics[clip,angle=-90,width=1\columnwidth]{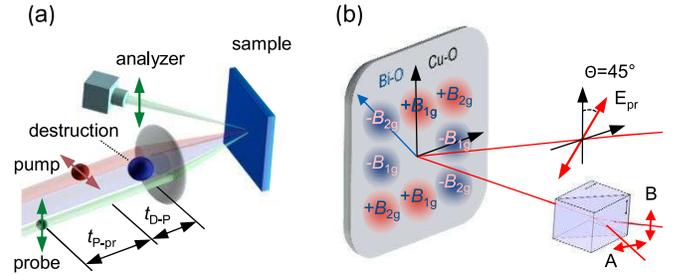}

\caption{(a) Schematic representation of the three-pulse experiment and notation
of delays between pulses. (b) Schematic representation of the detection
of $\Delta R_{\mathrm{B}{}_{\mathrm{1g}}}$ using a Wollaston prism.
The initial probe polarization is set at 45$^{\circ}$with respect
to the Cu-O bond direction. The intensity in channels A and B is balanced
and the differential signal is recorded, so that all contributions
except $\Delta R_{\mathrm{B}{}_{\mathrm{1g}}}$ are canceled out.\label{fig:schematics}}
\end{figure}
\begin{figure*}
\includegraphics[width=0.8\textwidth]{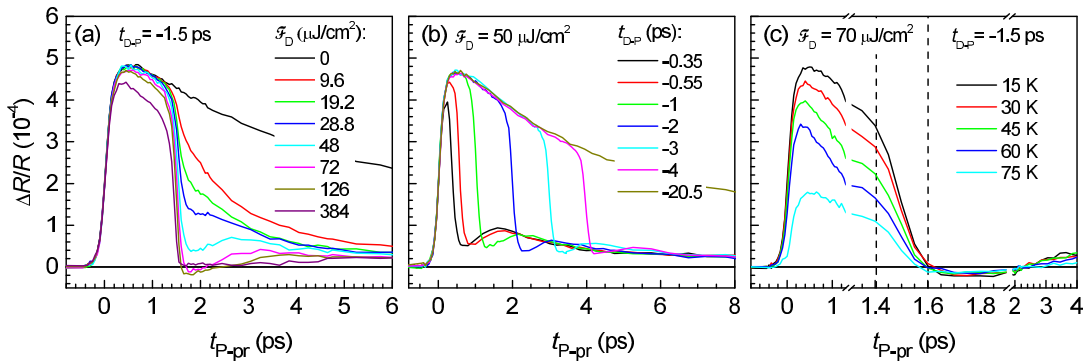}\caption{The influence of the destruction pulse on the transient reflectivity
in the superconducting state, when the D pulse arrives after the P
pulse using the PDS. (a) and (b) show dependence on the D-pulse fluence
and $t\mathrm{_{D-P}}$ delay at $T=15$ K, respectively. (c) $T$-dependence
of the transient reflectivity suppression.\label{fig:fig-destruction}}
\end{figure*}

While the all-optical transient response in the cuprates under weak
excitation can be well described in terms of the photoinduced absorption
of the photoexcited quasiparticles\cite{KabanovDemsar1999} the response
function in highly nonequilibrium states is unclear due to unknown
relative contributions of collective and single-particle degrees of
freedom to the transient optical reflectivity. To overcome this problem
the standard two-pulse all-optical pump-probe technique was extended
to a multi-pulse technique, which was shown to be instrumental in
extracting the order parameter dynamics in a charge density wave compound\cite{Yusupov2010}
as well as in the prototypical cuprate superconductor La$_{1.9}$Sr$_{0.1}$CuO$_{4}$\cite{Madan2016}. 

Here we extend our previous study\cite{Madan2016} of an ultrafast
S$\rightarrow$N$\rightarrow$S transition in La$_{1.9}$Sr$_{0.1}$CuO$_{4}$
to Bi$_{2}$Sr$_{2}$CaCu$_{2}$O$_{8+\delta}$ in search of universality,
and also to uncover potential imporant differences in the two materials
with substantially different critical temperatures and pseudogap/SC
gap ratios. By means of the all-optical multi-pulse technique combined
with the polarization selective optical probe we were able to separate
the SC state recovery dynamics from the previously studied\cite{Madan2015}
PG state recovery dynamic and enable discrimination between relaxation
in the nodal and anti-nodal BZ regions. The material has been studied
previously by time-resolved techniques\cite{giannetti2009discontinuity,CortesRettig2011,toda2011quasiparticle,SmallwoodHinton2012,SmallwoodZhang2014,SamallwoodZhang2015},
but thus far there has been no systematic study of the the non-equilibrium
transitions in this material, especially by the 3-pulse technique.

While we found that in La$_{1.9}$Sr$_{0.1}$CuO$_{4}$ the time dependent
Ginzburg-Landau (TDGL) theory can provide a fair quantitative description
of the SC order parameter recovery, only a qualitative description
of the data is possible in Bi2212, which we attribute to the large
SC order fluctuations in the PG state near time of the transition.
In addition, when only a partial SC state suppression is achieved,
the polarization resolved optical probe enables us to detect anisotropic
SC-order recovery timescales, revealing a faster SC gap recovery in
the anti-nodal direction in comparison with the nodal BZ regions.

\section{Experimental}

The sample used in this work was underdoped Bi$_{2}$Sr$_{2}$CaCu$_{2}$O$_{8+\delta}$
(Bi2212) single crystal with $T_{\mathrm{c}}\approx$ 78 K ($\delta=0.14$)
grown by means of the traveling solvent floating zone method. Before
mounting into a liquid-He flow cryostat the sample was freshly cleaved
using sticky tape.

The pulse train from a 250-KHz 1.55-eV Ti:Sapphire regenerative amplifier
was split into 50 fs destruction (D), pump (P) and probe (pr) pulse
trains that were independently delayed with respect to each other.
The resulting beams were focused and overlapped on the sample {[}see
Fig. \ref{fig:schematics} (a){]}. As in the standard pump-probe stroboscopic
experiments the transient reflectivity $\Delta R/R$ was measured
by monitoring the intensity of the weakest pr beam. The direct contribution
of the unchopped D beam to $\Delta R$ was rejected by means of a
lock-in synchronized to the chopper that modulated the intensity of
the P beam only. The fluences $\mathcal{F_{\mathrm{P}}}<5$ $\mu$J/cm$^{2}$
and $\mathcal{F_{\mathrm{pr}}}<3$ $\mu$J/cm$^{2}$ of the P and
pr pulses were kept in the linear response region, well below the
superconductivity destruction threshold\cite{giannetti2009discontinuity,stojchevska2011mechanisms,toda2011quasiparticle}.

To select different components of the anisotropic transient reflectivity\cite{Toda2014}
two different polarization sensitive detection schemes were used.
In the parallel detection scheme (PDS), which is sensitive to the
sum of $\Delta R_{A_{\mathrm{1g}}}$ and $\Delta R_{B_{\mathrm{1g}}}$
components\footnote{As in Ref. \onlinecite{Toda2014} we use the approximate notation
corresponding to the tetragonal symmetry.}, $\Delta R=\Delta R_{A_{\mathrm{g}}}+\Delta R_{B_{\mathrm{1g}}}$,
we used a single photodiode detection with an analyzer parallel to
the pr-beam polarization, where the pr-beam polarization was parallel
to the Cu-O bond direction. The polarizations of the P and D beams
were perpendicular to the pr-beam polarization in order to suppress
the signal due to P beam scattering.

In the balanced detection scheme (BDS), which is sensitive to the
$\Delta R_{B_{\mathrm{1g}}}$ component only, the pr-beam polarization
was oriented at 45\textdegree{} with respect to the Cu-O bond directions
and two photodiodes in combination with a Wollaston prism were used
for detection {[}see Fig. \ref{fig:schematics} (b){]}. When the polarization
axes of such detector are oriented along the Cu-O bond directions,
the difference of the two photodiode photocurrents corresponds to
$\Delta R_{B_{\mathrm{1g}}}$, while $\Delta R_{A_{\mathrm{1g}}}$
and $\Delta R_{B_{\mathrm{2g}}}$ components are rejected. Fine alignment
of the polarization and detector angles was done in the PG state at
120 K, to achieve a complete cancellation of the transient PG response. 

In order to suppress the P beam scattering contribution to the signal
in the BDS the P-beam frequency was doubled (3.1-eV P-photon energy)
and a long-pass filter in front of the detector was used while the
1.55-eV D-photon energy was the same as in the first scheme.\footnote{The scattering from the D-beam does not contribute significantly since
the beam is not modulated.}

\begin{figure*}
\includegraphics[angle=-90,width=0.8\textwidth]{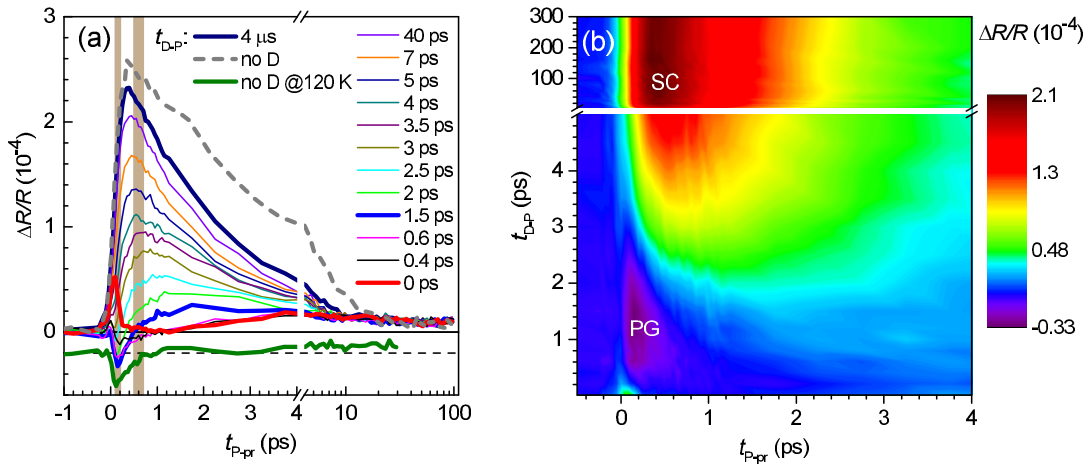}\caption{(a) Transient reflectivity using the PDS at $T=40$ K with $\mathcal{F}_{\mathrm{D}}=68$
$\mu$J/cm$^{2}$. For comparison a transient measured in the PG state
($T=120$ K) in the absence of the D pulse is shown vertically shifted
below the main data. The vertical shaded areas indicates the read-out
$t_{\mathrm{P-pr}}$ delays (see text). (b) The same data set as in
(a) shown as a colormap. At $t_{\mathrm{D-P}}=0$ both the PG and
SC signal are suppressed. With increasing $t_{\mathrm{D-P}}$ one
can observe a sequential recovery of the negative PG response followed
by the positive SC response.\label{fig:ParallelDetSch}}
\end{figure*}
\begin{figure*}
\includegraphics[width=0.8\textwidth]{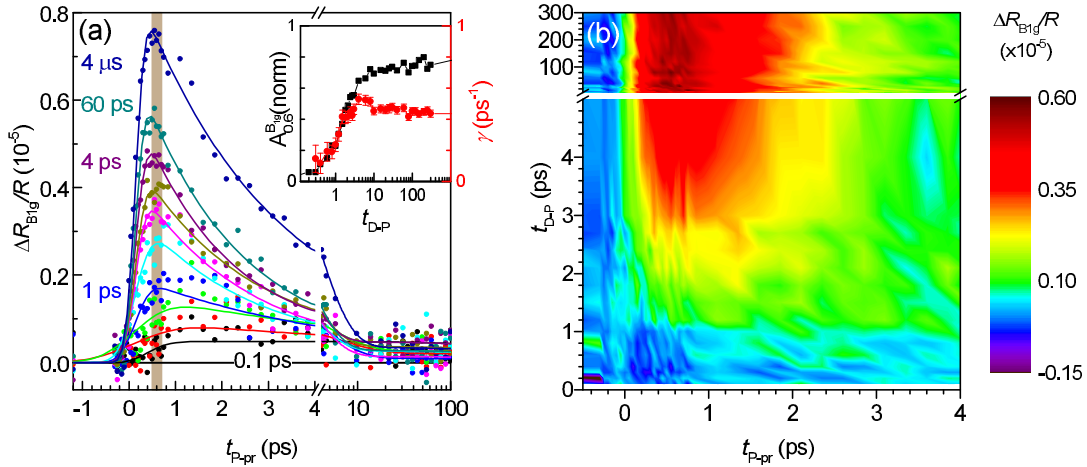}\caption{(a) Recovery of the transient-reflectivity $\mathrm{B}{}_{\mathrm{1g}}$
component at $T=40$ K and $\mathcal{F}_{\mathrm{D}}=56$ $\mu$J/cm$^{2}$.
The vertical shaded area represents the interval used to determine
$A_{\mathrm{SC}}$. Inset: black squares (left axis) - normalized
amplitude of the response, red circles - relaxation rate extracted
form single-exponential fits to the traces. Error bars are the standard
errors of the regression analysis. (b) The same data as in (a) shown
as a color map in $t_{\mathrm{D-P}}-t_{\mathrm{P-pr}}$. In the absence
of the PG response recovery of the SC signal is evident already at
$t_{\mathrm{D-P}}\sim1$ ps.\label{fig:BalancedDetSch}}
\end{figure*}

\section{Results}

\subsection{SC state destruction\label{sub:Destruction}}

To illustrate the destruction of the SC state, in Fig. \ref{fig:fig-destruction}
we plot the transient reflectivity for the case when the D pulse arrives
after the P pulse using the PDS. Depending on the D pulse fluence,
$\mathcal{F_{\mathrm{D}}}$, the transient reflectivity is suppressed
to different degrees. Above $\mathcal{F_{\mathrm{D}}\sim}70$ $\mu$J/cm$^{2}$,
SC order is completely suppressed on a 200-fs timescale after the
D-pulse arrival. Above $\mathcal{F_{\mathrm{D}}\sim}70$ $\mu$J/cm$^{2}$
we observe also a small negative overshot lasting a few hundred femtoseconds
followed by a weak recovery of the signal on a picosecond timescale.
Both features vanish at the highest fluence of $\sim400$ $\mu$J/cm$^{2}$.
The suppression timescale does not depend on the D-pulse arrival time
{[}Fig. \ref{fig:fig-destruction} (b){]} nor temperature {[}Fig.
\ref{fig:fig-destruction} (c){]}.
\begin{figure*}
\includegraphics[width=0.8\textwidth]{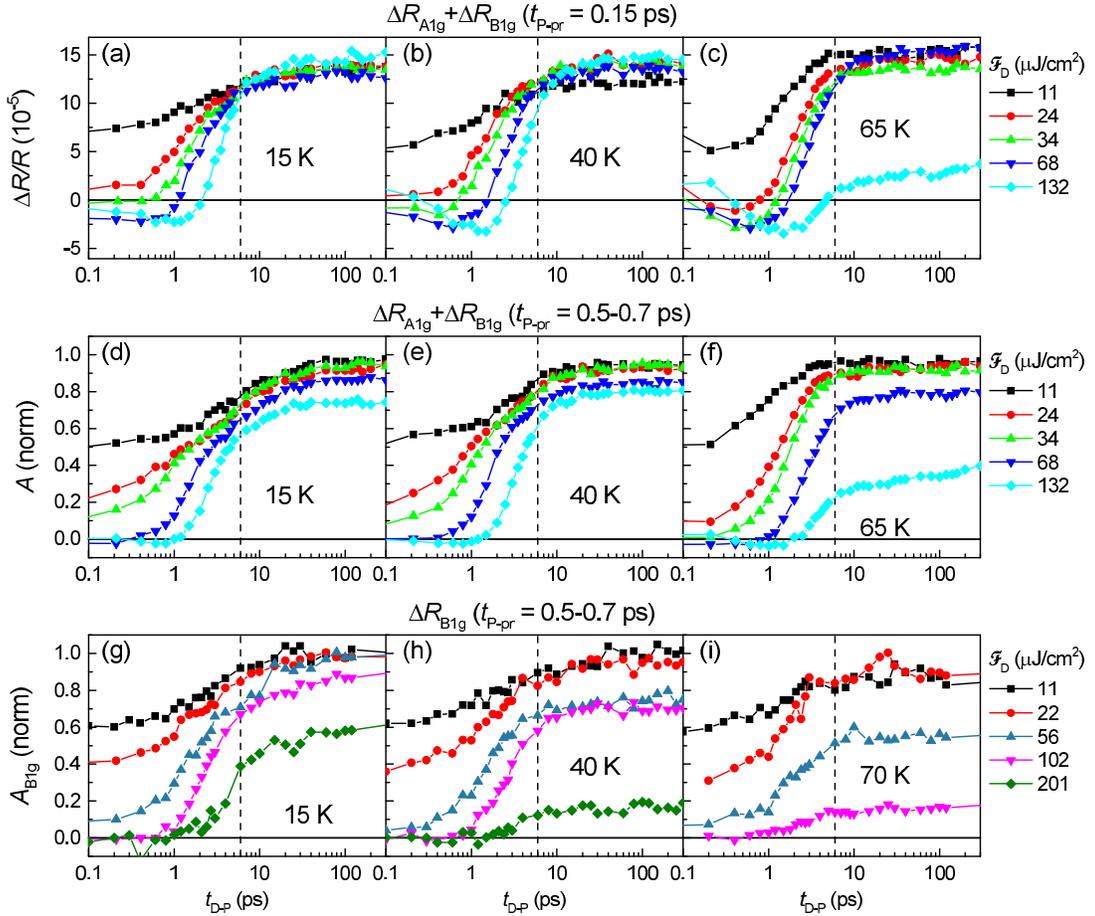}\caption{(a)-(c) The transient reflectivity at $t_{\mathrm{P-pr}}=0.15$ ps,
corresponding to the delay at which the PG response peaks, as a function
of $t_{\mathrm{D-P}}$ for different D-pulse fluences at different
temperatures. (d)-(f) Evolution of the normalized $\Delta R/R$ amplitude
averaged in the $0.5\mathrm{\,ps}\le t_{\mathrm{P-pr}}\le0.7\mathrm{\,ps}$
range as a function of $t_{\mathrm{D-P}}$ for different D-pulse fluences
at different temperatures. (g)-(i) the same for $\Delta R_{\mathrm{B}{}_{\mathrm{1g}}}/R$.
\label{fig:Trajectories}}
\end{figure*}

\subsection{SC state recovery}

In Fig. \ref{fig:ParallelDetSch} we show a typical transient reflectivity
data set measured in the PDS for the case when the P pulse arrives
after the D pulse. After a complete suppression for $t_{\mathrm{D-P}}\lesssim0.5$
ps we first observe a recovery of the negative PG component on a $1\sim$ps
timescale followed by the recovery of the positive SC component.

As shown previously\cite{Toda2014} the PG response does not contribute
to $\Delta R_{\mathrm{B}{}_{1g}}$ so recovery of the SC component
on the short $t_{\mathrm{D-P}}$ timescale can be observed more clearly
in the BDS. In Fig.$\,$\ref{fig:BalancedDetSch} we show a typical
transient reflectivity data set measured using the BDS. As expected,
the PG component is suppressed, but the signal-to-noise ratio is reduced
due to a smaller $\Delta R_{\mathrm{B}{}_{1g}}$ amplitude.

\section{Analysis and discussion}

\subsection{SC state destruction}

The destruction timescale of $\sim200$ fs is $T$ and $\mathcal{F}$
independent and faster than $\sim700$ fs in LSCO\cite{Kusar2008,BeyerStaedter2011}.
In LSCO it was suggested\cite{Kusar2008,BeyerStaedter2011} that the
high energy optical phonons created during the relaxation of the primary
photo electron-hole pair are the dominating pair breaking excitation
setting the destruction timescale. 

The faster destruction timescale in Bi2212 does not exclude the same
phonon mediated destruction mechanism since one polar optical phonon
can be generated by a photoexcited electron/hole every $\sim5$ fs\cite{BeyerStaedter2011}.
Taking the initial photo electron/hole energy of $\sim1$ eV and optical
phonon energy of 50 meV leads to $\sim100$ fs photo electron/hole
energy relaxation time that is fast enough to be compatible with the
experimental data\cite{Perfetti2007}. The phonon dominated pair-breaking
destruction of the SC state is supported also by the large optical
SC state destruction energy that is 5 times larger than the SC condensation
energy.\cite{stojchevska2011mechanisms}

\subsection{Analysis of the SC state recovery}

To analyze the recovery we first fit a finite-rise-time single-exponential
relaxation model to the transient reflectivity in Fig. \ref{fig:BalancedDetSch}
to obtain the $t_{\mathrm{D-P}}$ dependent relaxation rate $\gamma$.
In the inset to Fig. \ref{fig:BalancedDetSch} (a) we compare the
relaxation rate $\gamma$ from the fit to the amplitude of the $\mathrm{B}_{1g}$
SC response $A_{\mathrm{B1g}}=\overline{\Delta R_{\mathrm{B}_{1g}}}/\overline{\Delta R_{\mathrm{B}_{1g},\mathrm{no-D}}}$,
where $\overline{y}$ corresponds to the average of $y$ in the interval\footnote{In the vicinity of the peak of the unperturbed $\Delta R_{\mathrm{B_{1g}}}/R$-response}
$t_{\mathrm{P-pr}}=0.5-0.7$ ps and $\Delta R_{\mathrm{B}_{1g},\mathrm{no-D}}$
to the transient reflectivity in the absence of the D pulse. 

$\gamma$ and $A_{\mathrm{B1g}}$ initially recover on a similar time-scale
of $\sim4$ ps followed by slower dynamics extending towards the nanosecond
timescale. As in the case of (La,Sr)CuO$_{4+\delta}$ (LSCO), we attribute
the suppression of $\gamma$ during the first part of the recovery
to the critical slowing down of the SC fluctuations in the vicinity
of the transition.\cite{Madan2016} Upon the initial increase $\gamma$
decreases on the nanosecond timescale indicating cooling of the probed
volume: Since the effective temperature on longer timescales is far
from the critical temperature the $T$-dependence of $\gamma$ is
no longer critical but determined by the Rothwarf-Taylor bottleneck
dynamics.\cite{Kabanov2005}

Contrary to LSCO, where the PG component shows no suppression up to
a rather high excitation fluence,\cite{kuvsar2011dynamical} the PG
component in Bi2212 shows suppression already below\cite{Madan2015}
$\sim100$ $\mu$J/cm$^{2}$ so also the PG component is affected
by the D pulse. To extract the SC component recovery dynamics in the
PDS it is therefore necessary to take the PG dynamics into account. 

The PG component peaks at $t_{\mathrm{D-pr}}=0.15$ ps. Traces of
$\Delta R/R$ at this $t_{\mathrm{D-pr}}$ as function of $t_{\mathrm{D-P}}$
are shown in Fig. \ref{fig:Trajectories} (a-c). It is evident that
at higher $\mathcal{F}_{\mathrm{D}}$ the PG recovery leads to non-monotonous
traces due to the sub-ps recovery timescale\cite{Madan2015} of the
negative PG component preceding the recovery of the positive SC-state
component. Due to the rather fast PG-component relaxation time\cite{Toda2014}
{[}see Fig. \ref{fig:ParallelDetSch} (a){]} the contribution of the
PG component to $\Delta R/R$ should diminish with increasing $t_{\mathrm{D-pr}}$.
Taking $t_{\mathrm{P-pr}}$ in the interval $0.5\,\mathrm{ps}<t_{\mathrm{P-pr}}<0.7\,\mathrm{ps}$
where $\Delta R_{\mathrm{B_{1g}}}/R$ has a peak\footnote{The $\mathrm{A}_{1g}+\mathrm{B}_{1g}$ $\Delta R/R$ has a peak at
a slightly earlier time, where the PG component contribution is significantly
larger.} in the absence of the D pulse and the PG response is already suppressed,
we calculate the normalized average, $A=\overline{\Delta R}/\overline{\Delta R_{\mathrm{no-D}}}$.
Indeed, $A(t_{\mathrm{D-P}})$ traces presented in Fig. \ref{fig:Trajectories}
(d-f) show significantly less PG-component recovery and appear very
similar to the equivalent $A_{\mathrm{B1g}}(t_{\mathrm{D-P}})$ traces
shown in Fig. \ref{fig:Trajectories} (g-i).

\begin{figure}
\includegraphics[angle=-90,width=1\columnwidth]{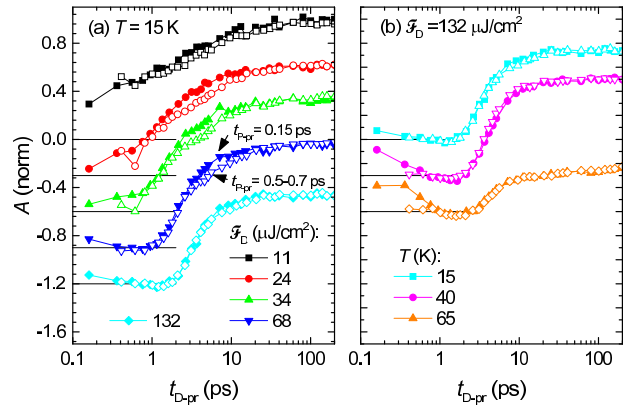}\caption{Comparison of $A\mathrm{_{S}}$ at two different $t_{\mathrm{P-pr}}$
as a function of $t_{\mathrm{D-pr}}$. The traces are vertically shifted
for clarity as indicated by the horizontal thin lines. Full and open
symbols correspond to $t_{\mathrm{P-pr}}=0.15$ ps and $t_{\mathrm{P-pr}}=0.5-0.7$
ps, respectively. The strongly PG-affected ($t_{\mathrm{P-pr}}=0.15$
ps) traces (full symbols) are vertically shifted and scaled to achieve
the best match for $t_{\mathrm{D-pr}}\gtrsim1$ ps.\label{fig:readout-comparison}}
\end{figure}

At $T>T_{\mathrm{c}}$ the PG recovers on the $\sim0.7$ ps timescale\cite{Madan2015}.
To check whether the amplitude of the PG component is modified during
the slower SC state recovery\cite{CoslovichGianetti2013} we compare
in Fig. \ref{fig:readout-comparison} the two readouts with different
PG contribution taken at $t_{\mathrm{P-pr}}=0.15$ ps and the average
in the interval $0.5\,\mathrm{ps}<t_{\mathrm{P-pr}}<0.7\,\mathrm{ps}$
($A$). At the highest $\mathcal{F}_{\mathrm{D}}=132$ $\mu$J/cm$^{2}$
it is possible to overlap the traces beyond $t_{\mathrm{D-pr}}\gtrsim1$
ps at all measured temperatures when plot as a function of\footnote{The difference in sampling time of 0.55 ps needs to be taken into
account when directly comparing the traces. } $t_{\mathrm{D-pr}}$ by vertically shifting\footnote{The fully recovered PG component contributes to a $t_{\mathrm{D-P}}$-independent
negative shift at longer $t_{\mathrm{D-P}}$. } and rescaling. At intermediate $\mathcal{F}_{\mathrm{D}}$s the complete
overlap is not possible. The shifted and rescaled readouts at $t_{\mathrm{P-pr}}=0.15$
ps show slightly higher values in the $\sim2$ - $\sim10$-ps delay
range. This could indicate that the negative PG response at 1.55-eV
probe-photon energy is transiently suppressed\footnote{When the negative PG component is suppressed the total $\Delta R/R$
increases.} by the appearance of the SC order. 

A possibly related suppression of the PG component in the SC state
at 1.08-eV probe-photon energy was suggested recently\cite{CoslovichGianetti2013}.
Considering an earlier report\cite{toda2011quasiparticle}, however,
where by selecting a particular polarization and probe-photon energy
no suppression of the PG component in the SC state was observed, we
attribute the difference between readouts in our experiment to the
SC-gap dependent pre-bottleneck SC-state dynamics, which influences
the readouts at $t_{\mathrm{D-P}}=0.15$ ps.

\subsection{SC state recovery timescale in nodal and anti-nodal response\label{sub:Recovery-timescale}}

\begin{figure}
\includegraphics[width=0.95\columnwidth]{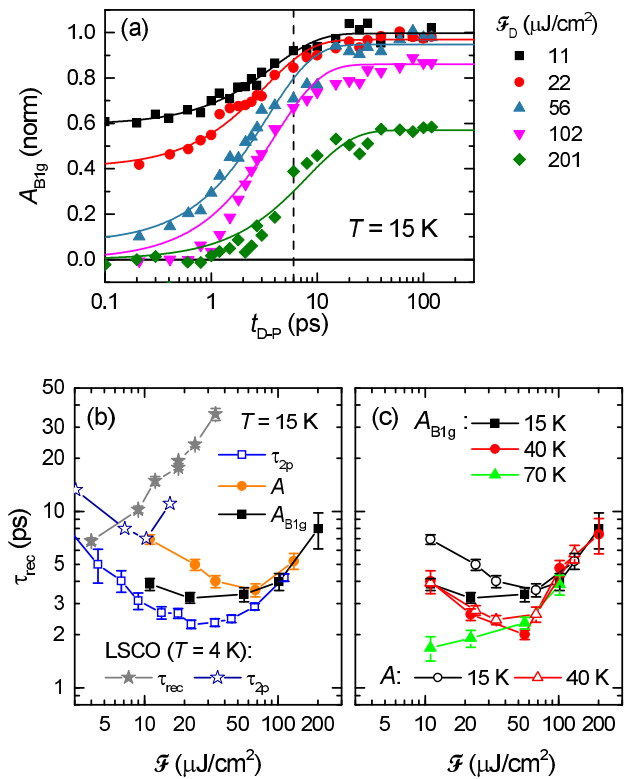}\caption{(a) Fits of Eq. (\ref{eq:ExpRec}) to the $A_{\mathrm{B1\mathrm{g}}}$
trajectories at $T=15$ K. (b) The recovery time of the superconducting
response from the fits at $T=15$ K as a function of fluence for $A_{\mathrm{B1\mathrm{g}}}$
and $A$ trajectories {[}Fig. \ref{fig:Trajectories} (d) and (g){]}.
For comparison the $\Delta R/R$ relaxation time from a two-pulse
experiment at 15 K in Bi2212 is shown by open squares. The corresponding
relaxation times in LSCO\cite{kusthesis} at $T=4$ K are shown by
stars. (c) Temperature dependence of $\tau_{\mathrm{rec}}$ of $A_{\mathrm{B1\mathrm{g}}}$
(full symbols) and $A$ (open symbols) trajectories. \label{fig:Recovery-time}}
\end{figure}

In Fig.$\,$\ref{fig:Recovery-time} (b) we compare the fluence dependencies
of the SC recovery time, $\tau_{\mathrm{rec}}$, for both symmetries
to the standard 2-pulse transient-reflectivity relaxation time, $\tau_{\mathrm{2p}}$.
We estimate $\tau_{\mathrm{rec}}$ using a phenomenological exponential
fit:
\begin{equation}
A\mathrm{_{S}}=A_{\mathrm{T}}-A_{\mathrm{e}}e^{-t_{\mathrm{D-P}}/\tau_{\mathrm{rec}}},\label{eq:ExpRec}
\end{equation}
to the trajectories in Fig.$\,$\ref{fig:Trajectories}(d)-(i). The
lower $\mathcal{F}_{\mathrm{D}}$ data can be rather well fit using
the simple single-exponential recovery model (\ref{eq:ExpRec}) while
at higher $\mathcal{F}_{\mathrm{D}}$ the recovery appears clearly
non exponential as shown in Fig \ref{fig:Recovery-time} (a). 

As a function of fluence both, $\tau_{\mathrm{rec}}$ and $\tau_{\mathrm{2p}}$
show a minimum at intermediate fluences. Above $\mathcal{F}_{\mathrm{D}}\sim$100
$\mu$J/cm$^{2}$ timescales of different signals match rather well\footnote{Despite the worse fit quality. }
and show virtually no $T$-dependence {[}see Fig. \ref{fig:Recovery-time}
(c){]}. At low $\mathcal{F}_{\mathrm{D}}$ and $T=15$ K, however,
the A$_{1\mathrm{g}}$-dominated-channel $\tau_{\mathrm{rec}}$ slows
down much more with decreasing $\mathcal{F}_{\mathrm{D}}$ than the
B$_{\mathrm{1g}}$-channel $\tau_{\mathrm{rec}}$ and $\tau_{\mathrm{2p}}$.
On the contrary, at $T=40$ K, both, the A$_{1\mathrm{g}}$-dominated-
and B$_{\mathrm{1g}}$-channel $\tau_{\mathrm{rec}}$ show identical
fluence dependence in the full $\mathcal{F}_{\mathrm{D}}$ range with
a sharp up-turn at $\mathcal{F}_{\mathrm{D}}\sim60$ $\mu$J/cm$^{2}$. 

A faster $\Delta R/R$ relaxation in the B$_{1\mathrm{g}}$ configuration
has been observed already in the low excitation 2-pulse experiments.\cite{Toda2014}
Due to the sensitivity of the the B$_{1\mathrm{g}}$ configuration
to the anti-nodal Brillouin-zone (BZ) region this is consistent with
a faster quasiparticle relaxation around anti-nodes\footnote{An consistent increase of the QP relaxation time away from the nodal
point was observed in a recent ARPES experiment\cite{SamallwoodZhang2015}
in some controversy to an earlier ARPES result\cite{CortesRettig2011}.} either by recombination or by scattering into the nodal BZ region,
which contributes to the A$_{1\mathrm{g}}$-dominated channel showing
a slower decay. 

The effect of the faster antinodal quasiparticle relaxation is also
evident in our 3-pulse experiment, but only, when the SC order is
not completely suppressed and $T$ is below $\sim40$ K. From the
3-pulse data it appears that upon a modest suppression the SC gap
recovers faster at the anti-nodes than near the nodes. 

At higher $\mathcal{F}_{\mathrm{D}}$ upon a complete suppression
of the SC gap our data suggest the recovery that is more homogeneous
across the Fermi surface. This could be attributed to two factors.
First, during the initial part of the recovery the suppression of
the Rothwarf-Taylor phonon bottleneck\cite{Kabanov2005} and lifting
of the SC-gap-imposed QP-relaxation phase space restrictions\cite{SmallwoodMiller2016}
enable efficient transfer of the excess QP energy to the phonon bath
together with efficient diffusion of excitations across all of the
BZ. Second, at higher $\mathcal{F}_{\mathrm{D}}$ the lattice bath
is heated closer to $T_{\mathrm{c}}$ so the QP-relaxation phase space
restrictions can be easier overcome by the phonon assisted QP scattering.

In LSCO {[}Fig. \ref{fig:Recovery-time} (b){]} $\tau_{\mathrm{rec}}$
is similarly to $\tau_{\mathrm{2p}}$ significantly longer than in
Bi2212.\cite{toda2011quasiparticle} The generally slower $\tau_{\mathrm{2p}}$
and $\tau_{\mathrm{rec}}$ in LSCO could be attributed to the smaller
SC gap enhancing the quasiparticle relaxation bottleneck\cite{KabanovDemsar1999}.
Moreover, in LSCO $\tau_{\mathrm{rec}}$ increases monotonically above
the destruction threshold fluence,\cite{Kusar2008} $\mathcal{F}_{\mathrm{Dth}}=4.2$
$\mu$J/cm$^{2}$, while in Bi2212 with\cite{toda2011quasiparticle}
$\mathcal{F}_{\mathrm{Dth}}=\mbox{\ensuremath{\sim}16}$ $\mu$J/cm$^{2}$,
the increase starts only above $\sim4\mathcal{F}_{\mathrm{Dth}}$.
This could be attributed to the lattice temperature after the quench
being closer\footnote{In LSCO the lattice temperature after the destruction pulse reaches\cite{Madan2016}
$T_{\mathrm{c}}$ at $\mathcal{F}_{\mathrm{D}}=20$$\mu$J/cm$^{2}$.} to $T_{\mathrm{c}}$ in LSCO than in Bi2212 resulting in a stronger
critical slowing down of the SC order parameter dynamics.

\subsection{Time dependent Ginzburg-Landau model}

We proceed by analyzing the trajectory of the SC amplitude through
the transition in the framework of the time-dependent Ginzburg-Landau
(TDGL) theory. In previous study\cite{Madan2016} of the SC-order
recovery in LSCO we have shown that the TDGL theory fails to consistently
describe the ultrafast optical destruction of the SC condensate. On
the other hand,\emph{ the SC condensate recovery can be quantitatively
modeled }using a phenomenological response function and the Ginzburg-Landau
time, $\tau_{\mathrm{GL}}$, as the only free fit parameter assuming
a finite magnitude of the initial depth-dependent order parameter
(Fig. 3c in Ref. {[}\onlinecite{Madan2016}{]}). The magnitude of
the initial order parameter corresponds to the magnitude of the frozen
SC fluctuations after the quench from the normal/PG to the SC state
which is a function of the depth-dependent quench-rate (Eq. (4) in
Ref. {[}\onlinecite{Madan2016}{]}). In LSCO even better fit is possible
using a phenomenological depth-dependent initial order parameter $\psi_{\mathrm{BC}}(z)$:
\begin{equation}
\psi_{\mathrm{BC}}(z)=\begin{cases}
c\,z & ;\,\mathit{U\mathcal{\mathrm{_{D}}}}(z)>U{}_{\mathrm{th}}\\
\sqrt{1-T/T_{\mathrm{c}}} & ;\,\mathit{U\mathcal{\mathrm{_{D}}}}(z)\leq U{}_{\mathrm{th}},
\end{cases}\label{eq:IC}
\end{equation}
where $c$ is an additional $\mathcal{F_{\mathrm{D}}}$-dependent
free parameter, $\mathit{U\mathcal{\mathrm{_{D}}}}(z)$ the depth-dependent
absorbed optical-energy density and $U{}_{\mathrm{th}}$ the SC-destruction-threshold
optical-energy density. 

In the following we apply a similar TDGL approach to the SC state
recovery dynamics in Bi2212.

\subsubsection{Response function}

As a starting point one needs to establish the relation between the
superconducting order parameter magnitude, $\left|\psi_{\mathrm{GL}}\right|$,
and the transient optical response amplitude. This relation was in
the case of LSCO established phenomenologically from the temperature
dependence of the normalized weak-excitation $\Delta R/R$ amplitude,
$A_{\mathrm{S}}$. 

\begin{figure}
\includegraphics[width=1\columnwidth]{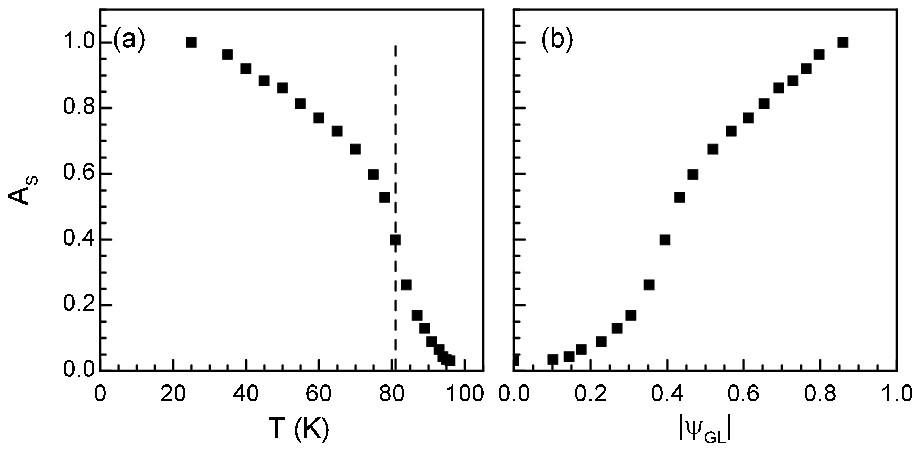}\medskip{}

\includegraphics[width=0.98\columnwidth]{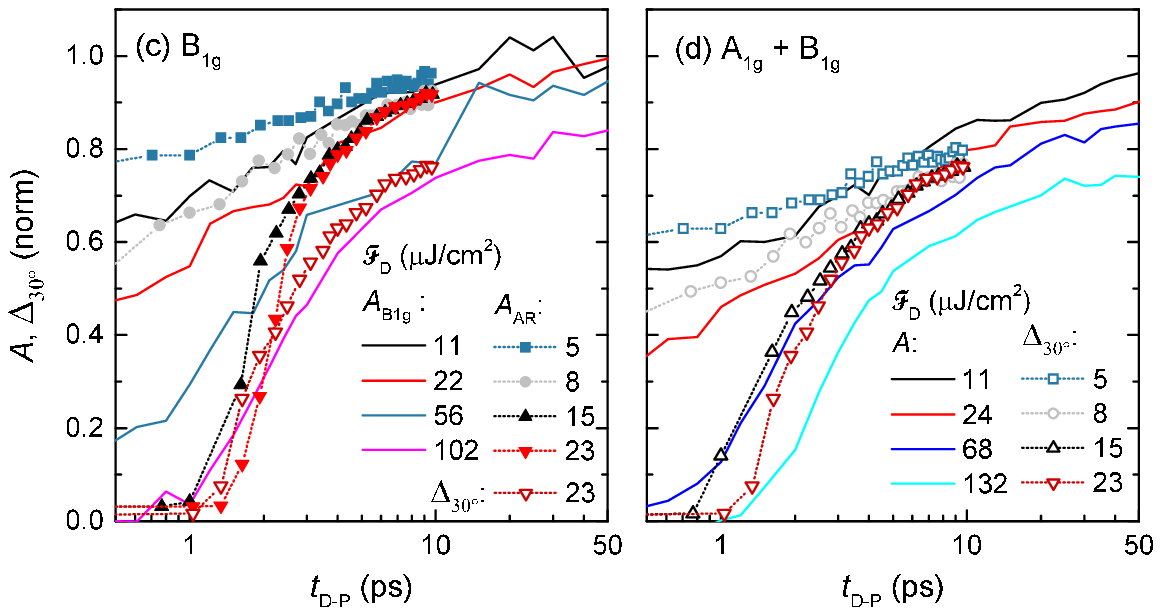}\caption{(a) The amplitude of the normalized transient superconducting response
as a function of temperature. The dashed vertical line marks the critical
temperature measured by a SQUID magnetometer. (b) The amplitude of
the the normalized transient superconducting response as a phenomenological
function of the GL order parameter obtained from (a) assuming a mean-field
$T_{\mathrm{c}}^{\mathrm{MF}}=96$ K. This relation is used as the
response function for the theoretical calculation of the superconducting
order parameter evolution presented in Fig. \ref{fig:Theory}. (c)
Comparison of $A\mathrm{_{B_{1g}}}$ to $A{}_{\mathrm{AR}}^{\mathrm{}}$
obtained form the TR-ARPES gap dynamics\cite{SmallwoodZhang2014}
using the response function from (b). \label{fig:ARPES-comparison}
The normalized TR-ARPES gap (open symbols) at the highest $\mathcal{F}_{\mathrm{ARPES}}$
is also shown for comparison. (d) Comparison of $A\mathrm{_{0.6}}$
to to the normalized TR-ARPES gap. }
\end{figure}

In Bi2212 $A_{\mathrm{S}}$ does not go to zero at $T_{\mathrm{c}}$
due to the large pairing fluctuations\cite{ZhangSmallwood2013,madan2014separating}
above $T_{\mathrm{c}}$ as shown in Fig.$\,$\ref{fig:ARPES-comparison}
(a), inconsistently with the the GL theory. However, by taking into
account that $A_{\mathrm{S}}$ is sensitive to the paring amplitude\cite{madan2014separating}
and not the SC phase coherence, we can still apply the GL description
assuming that only the SC phase coherence is established at $T{}_{\mathrm{c}}$,
while the largest temperature at which $A_{\mathrm{S}}$ is still
observable corresponds to the mean-field pairing critical temperature,
$T_{\mathrm{c}}^{\mathrm{MF}}\simeq96$ K. Implying the standard GL
$T$-dependence of the (pairing) order parameter, $\psi_{\mathrm{GL}}\propto\sqrt{1-T/T_{\mathrm{c}}^{\mathrm{MF}}}$,
we can now establish the response function as shown in Fig.$\,$\ref{fig:ARPES-comparison}(b).

To further analyze the link between the SC order parameter and $A_{\mathrm{S}}$
we compare our results to recent TR-ARPES gap dynamics data in near-optimally
doped Bi2212 ($T\mathrm{_{c}}=91$ K).\cite{SmallwoodZhang2014} Considering
the different doping levels of the samples and different spatial\footnote{The optical penetration depth in Bi2212\cite{stojchevska2011mechanisms}
is of the order of 100 nm in comparison to a nm scale photelectron
escape depth. } and reciprocal-space\footnote{The B$_{1\mathrm{g}}$ optical response is sensitive to a broad region
near the anti node while the A$_{1\mathrm{g}}$ response samples both
the nodal and antinodal regions\cite{DevereauxHackl07}} sensitivity of the probes only a qualitative correspondence between
the results is expected, since even in the case of TR-ARPES the extraction
of the SC gap is not rigorously defined\cite{SmallwoodZhang2014}.
For the sake of comparison we therefore assume that $\left|\psi_{\mathrm{GL}}(t)\right|\propto\Delta_{30{^\circ}}(t)$,
where $\Delta_{30{^\circ}}(t)$ corresponds to the TR-ARPES gap at
the edge of the Fermi arc and calculate $A_{\mathrm{\mathrm{AR}}}(t)$
using the response function in Fig.$\,$\ref{fig:ARPES-comparison}(b).
In Fig. \ref{fig:ARPES-comparison} (c) and (d) we show comparison
of our data to both, $A_{\mathrm{\mathrm{AR}}}$ and the TR-ARPES
gap, $\Delta_{30{^\circ}}$. 

For the B$_{1\mathrm{g}}$ configuration we find a surprisingly good
match between $A_{\mathrm{\mathrm{AR}}}$ and A$_{\mathrm{B}1\mathrm{g}}$
in the low fluence \newcounter{fnnumber}
\footnote{Due to the exponential decay of the excitation fluence away from the surface the equivalent external  in all-optical experiment is $\sim 1.5$ times larger than in the case of TR-ARPES.}
\setcounter{fnnumber}{\thefootnote} region, where the SC gap is only partially suppressed. At higher
$\mathcal{F}$, where the gap\footnote{The surface gap in the case of TR-ARPES and the bulk gap in the case
of optics.} is completely suppressed, the dynamics appears significantly different
below $\sim3$ ps, unless we compare curves with very different fluences.
Ignoring the response function a direct comparison of $\Delta_{30{^\circ}}$
at $\mathcal{F}=23$ $\mu$J/cm$^{2}$ to $A{}_{\mathrm{B}1\mathrm{g}}$
at 4.4 times higher $\mathcal{F}=102$ $\mu$J/cm$^{2}$ gives a good
match in the region of the strong suppression of the gap. 

For the $A{}_{1\mathrm{g}}$ dominated configuration a better match
is observed when we compare $A$ to $\Delta_{30{^\circ}}$ directly
{[}Fig. \ref{fig:ARPES-comparison} (d){]} while $A_{\mathrm{AR}}^{\mathrm{}}$
shows consistently higher magnitude than $A$. Similarly to the B$_{1\mathrm{g}}$
configuration, a good match is observed at a complete SC gap suppression
between the TR-ARPES trajectory at $\mathcal{F}=15$ $\mu$J/cm$^{2}$
and $A$ at 4.5 times higher $\mathcal{F}=68$ $\mu$J/cm$^{2}$. 

Assuming that the TR-ARPES SC gap dynamics \emph{is identical to the
bulk gap} dynamics the difference between the fluences of the corresponding-timescales
data can be, at least partially\footnotemark[\thefnnumber], attributed
to the smearing of the optical-probe dynamics due to the depth-dependent
excitation density and SC gap suppression. This is corroborated by
the convergence of the optical and TR-ARPES trajectories with similar
fluence on longer timescales, when the spatial inhomogeneity is expected
to decrease. 

The inaccuracy of the empirical response function {[}Fig. \ref{fig:ARPES-comparison}
(b){]} can further contribute to the difference, especially in the
region of small SC order parameter. Contrary to LSCO\cite{Madan2016}
where the response function is linear up to $A_{\mathrm{S}}\sim0.8$
the s-shape of the response function in the present case suggests
that $A_{\mathrm{AR}}^{\mathrm{}}$ might be underestimated for low
values of the gap.

Importantly, taking into account the inherent differences between
the techniques we can conclude that the TR-ARPES Fermi-arc SC gap
and the antinodal SC gap inferred from the B$_{\mathrm{1g}}$ channel
multi-pulse optical probe show qualitatively identical suppression
and recovery dynamics.

\subsubsection{Simulations}

As in the case of LSCO\cite{Madan2016} we simulate the evolution
of the order parameter through the transition by solving the dimensionless
form of the first of the two TDGL equations: 
\begin{equation}
\frac{\partial\psi}{\partial t}=\alpha_{r}(t,z)\psi-\psi|\psi|^{2}+\nabla^{2}\psi,\label{eq:TDGL}
\end{equation}
where time and length are measured in units of $\tau_{\mathrm{GL}}$
(fit parameter) and the coherence length, respectively. $\alpha_{r}(t,z)$
is a time- and depth-dependent reduced temperature which is the solution
of the three temperature model\cite{Perfetti2007} combined with the
heat diffusion equation.\cite{Madan2016}

We neglect the second TDGL equation and any lateral variation of the
order parameter, assuming that all the Kibble-Zurek (KZ) physics\cite{Kibble1997,Zurek1996}
can be phenomenologically absorbed into the initial order parameter
$\psi_{\mathrm{BC}}(z)$ using (\ref{eq:IC}), and the phase dynamics,
i.e. the dynamics of vortices, does not significantly modify the order
parameter amplitude. 

\begin{figure}
\includegraphics[width=1\columnwidth]{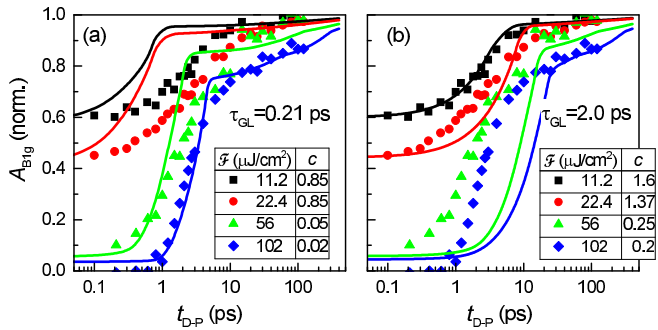}\caption{Comparison of the simulated amplitude of the transient superconducting
response to the B$_{1g}$ experimental data measured at 15 K. (a)
and (b) correspond to different values of $\tau_{\mathrm{GL}}$ indicated
in the graphs. Tables show the values of the parameter $c$ defining
the initial order parameter {[}Eq. (\ref{eq:IC}){]}.\label{fig:Theory}}
\end{figure}

The two fitting parameters $c$ and $\tau_{\mathrm{GL}}$ are rather
independent. While the first defines the Kible-Zurek-physics-related
amplitude of the response at $t=0$, the second defines the time-scale
of the recovery. In Fig.$\,$\ref{fig:Theory} we present typical
results of the simulations for two different values of the $\tau_{\mathrm{GL}}$
optimized to fit the highest-$\mathcal{F}_{\mathrm{D}}$ and the lowest-$\mathcal{F}_{\mathrm{D}}$
trajectories at 15 K, respectively. One can see that while a decent
agreement for a targeted curve can be achieved, one needs to significantly
vary $\tau_{\mathrm{GL}}$ to fit the complete data set. Since such
variation is unphysical, we can state that the presented TDGL approach
is only sufficient to describe the present data qualitatively, contrary
to what was found in LSCO, where a more quantitative description is
possible.

The lack of quantitative description can not be attributed to the
omission of the second TDGL equation and the resulting vortex dynamics
it describes. While at a partial SC-state order parameter suppression
no KZ-vortices formation is expected more vortices would be created
with further suppression. The presence of vortices at increased order
parameter suppression is expected to further slow down the SC-state
recovery\footnote{Supplemental information to Ref. {[}\onlinecite{Madan2016}{]} }.
Looking at Fig. \ref{fig:Theory} one can clearly see, that even without
the vortex dynamics the TDGL solutions display a stronger recovery-timescale
slowdown with increased order parameter suppression than the experimental
data, so inclusion of the vortex dynamics into modeling is expected
to only increase the discrepancy.

On the other hand, the lack of quantitative description is not very
surprising due to the large pairing fluctuations contribution\cite{madan2014separating}
to the transient reflectivity above $T_{\mathrm{c}}$ that prevents
strict applicability of the TDGL theory and undermine the phenomenological
link between the order parameter magnitude and the experimentally
observable $A_{\mathrm{S}}$.

\section{Summary an conclusions}

Our systematic investigation of the ultrafast optical suppression
and recovery of the superconducting state in Bi$_{2}$Sr$_{2}$CaCu$_{2}$O$_{8+\delta}$
by means of polarization-selective multi-pulse optical time-resolved
spectroscopy leads to some interesting, and somewhat surprising new
findings. We found that the SC order is suppressed on the 200-fs timescale,
comparable to the recent laser TR-ARPES\cite{SmallwoodZhang2014}
results. The destruction timescale is independent of the temperature
and optical destruction pulse energy and is consistent with a photoexcited
carrier energy-transfer to the high-energy pair breaking phonons.

The recovery of the SC order is slower appearing on the 2-8 ps timescale
showing non-monotonous dependence on the destruction pulse energy.
At low $T$ and a partial SC-state suppression the data shows that
the SC gap in the antinodal region recovers faster than near the nodes.
Perhaps surprisingly, the recovery also slows down with decreasing
$T$ highlighting the importance of thermal fluctuations in the recovery
mechanism. When the SC state is strongly suppressed, the recovery
becomes non-exponential with the recovery timescale slowing down,
becoming $T$-independent.

The fact that the antinodal SC order parameter recovery dynamics inferred
from the B$_{\mathrm{1g}}$ channel and the TR-ARPES Fermi-arc SC
gap dynamics\cite{SmallwoodZhang2014} show qualitatively identical
recovery dynamics gives us confidence in the significance of the multipulse
technique.

Despite strong SC fluctuations above $T_{\mathrm{c}}$ and the anisotropic
SC-gap recovery the time dependent Ginzburg-Landau model qualitatively
describes the SC-order temporal dynamics reasonably well, considering
its limitations.
\begin{acknowledgments}
The authors acknowledge the financial support of Slovenian Research
Agency (research core funding No-P1-0040) and European Research Council
Advanced Grant TRAJECTORY (GA 320602) for financial support. We would
like to thank L. Stojchevska for helping with the measurements.
\end{acknowledgments}

\bibliographystyle{apsrev4-1}
\bibliography{BSCCO}

\begin{thebibliography}{48}%
\makeatletter
\providecommand \@ifxundefined [1]{%
 \@ifx{#1\undefined}
}%
\providecommand \@ifnum [1]{%
 \ifnum #1\expandafter \@firstoftwo
 \else \expandafter \@secondoftwo
 \fi
}%
\providecommand \@ifx [1]{%
 \ifx #1\expandafter \@firstoftwo
 \else \expandafter \@secondoftwo
 \fi
}%
\providecommand \natexlab [1]{#1}%
\providecommand \enquote  [1]{``#1''}%
\providecommand \bibnamefont  [1]{#1}%
\providecommand \bibfnamefont [1]{#1}%
\providecommand \citenamefont [1]{#1}%
\providecommand \href@noop [0]{\@secondoftwo}%
\providecommand \href [0]{\begingroup \@sanitize@url \@href}%
\providecommand \@href[1]{\@@startlink{#1}\@@href}%
\providecommand \@@href[1]{\endgroup#1\@@endlink}%
\providecommand \@sanitize@url [0]{\catcode `\\12\catcode `\$12\catcode
  `\&12\catcode `\#12\catcode `\^12\catcode `\_12\catcode `\%12\relax}%
\providecommand \@@startlink[1]{}%
\providecommand \@@endlink[0]{}%
\providecommand \url  [0]{\begingroup\@sanitize@url \@url }%
\providecommand \@url [1]{\endgroup\@href {#1}{\urlprefix }}%
\providecommand \urlprefix  [0]{URL }%
\providecommand \Eprint [0]{\href }%
\providecommand \doibase [0]{http://dx.doi.org/}%
\providecommand \selectlanguage [0]{\@gobble}%
\providecommand \bibinfo  [0]{\@secondoftwo}%
\providecommand \bibfield  [0]{\@secondoftwo}%
\providecommand \translation [1]{[#1]}%
\providecommand \BibitemOpen [0]{}%
\providecommand \bibitemStop [0]{}%
\providecommand \bibitemNoStop [0]{.\EOS\space}%
\providecommand \EOS [0]{\spacefactor3000\relax}%
\providecommand \BibitemShut  [1]{\csname bibitem#1\endcsname}%
\let\auto@bib@innerbib\@empty
%</preamble>
\bibitem [{\citenamefont {Bunkov}\ and\ \citenamefont
  {Godfrin}(2000)}]{Bunkov2000}%
  \BibitemOpen
  \bibfield  {author} {\bibinfo {author} {\bibfnamefont {Y.~M.}\ \bibnamefont
  {Bunkov}}\ and\ \bibinfo {author} {\bibfnamefont {H.}~\bibnamefont
  {Godfrin}},\ }\href {\doibase 10.1007/978-94-011-4106-2} {\emph {\bibinfo
  {title} {{Topological Defects and the Non-Equilibrium Dynamics of Symmetry
  Breaking Phase Transitions}}}},\ edited by\ \bibinfo {editor} {\bibfnamefont
  {Y.~M.}\ \bibnamefont {Bunkov}}\ and\ \bibinfo {editor} {\bibfnamefont
  {H.}~\bibnamefont {Godfrin}}\ (\bibinfo  {publisher} {Springer Netherlands},\
  \bibinfo {address} {Dordrecht},\ \bibinfo {year} {2000})\ p.\ \bibinfo
  {pages} {396}\BibitemShut {NoStop}%
\bibitem [{\citenamefont {Higgs}(1966)}]{Higgs1966}%
  \BibitemOpen
  \bibfield  {author} {\bibinfo {author} {\bibfnamefont {P.~W.}\ \bibnamefont
  {Higgs}},\ }\href {\doibase 10.1103/PhysRev.145.1156} {\bibfield  {journal}
  {\bibinfo  {journal} {Physical Review}\ }\textbf {\bibinfo {volume} {145}},\
  \bibinfo {pages} {1156} (\bibinfo {year} {1966})}\BibitemShut {NoStop}%
\bibitem [{\citenamefont {Volovik}(2010)}]{Volovik2003}%
  \BibitemOpen
  \bibfield  {author} {\bibinfo {author} {\bibfnamefont {G.}~\bibnamefont
  {Volovik}},\ }\href {\doibase 10.1080/00107510903360394} {\emph {\bibinfo
  {title} {Contemporary Physics}}},\ \bibinfo {number} {5}\ (\bibinfo
  {publisher} {Oxford university press},\ \bibinfo {year} {2010})\ pp.\
  \bibinfo {pages} {451--452}\BibitemShut {NoStop}%
\bibitem [{Note1()}]{Note1}%
  \BibitemOpen
  \bibinfo {note} {The quasiparticle energy distribution is nonthermal so
  strictly speaking the electronic $T$ is not well defined.}\BibitemShut
  {Stop}%
\bibitem [{\citenamefont {Carnahan}\ \emph {et~al.}(2004)\citenamefont
  {Carnahan}, \citenamefont {Kaindl}, \citenamefont {Orenstein}, \citenamefont
  {Chemla}, \citenamefont {Oh},\ and\ \citenamefont
  {Eckstein}}]{CarnahanKaindl2004}%
  \BibitemOpen
  \bibfield  {author} {\bibinfo {author} {\bibfnamefont {M.}~\bibnamefont
  {Carnahan}}, \bibinfo {author} {\bibfnamefont {R.}~\bibnamefont {Kaindl}},
  \bibinfo {author} {\bibfnamefont {J.}~\bibnamefont {Orenstein}}, \bibinfo
  {author} {\bibfnamefont {D.}~\bibnamefont {Chemla}}, \bibinfo {author}
  {\bibfnamefont {S.}~\bibnamefont {Oh}}, \ and\ \bibinfo {author}
  {\bibfnamefont {J.}~\bibnamefont {Eckstein}},\ }\href@noop {} {\bibfield
  {journal} {\bibinfo  {journal} {Physica C: Superconductivity}\ }\textbf
  {\bibinfo {volume} {408}},\ \bibinfo {pages} {729} (\bibinfo {year}
  {2004})}\BibitemShut {NoStop}%
\bibitem [{\citenamefont {Kusar}\ \emph {et~al.}(2008)\citenamefont {Kusar},
  \citenamefont {Kabanov}, \citenamefont {Demsar}, \citenamefont {Mertelj},
  \citenamefont {Sugai},\ and\ \citenamefont {Mihailovic}}]{Kusar2008}%
  \BibitemOpen
  \bibfield  {author} {\bibinfo {author} {\bibfnamefont {P.}~\bibnamefont
  {Kusar}}, \bibinfo {author} {\bibfnamefont {V.~V.}\ \bibnamefont {Kabanov}},
  \bibinfo {author} {\bibfnamefont {J.}~\bibnamefont {Demsar}}, \bibinfo
  {author} {\bibfnamefont {T.}~\bibnamefont {Mertelj}}, \bibinfo {author}
  {\bibfnamefont {S.}~\bibnamefont {Sugai}}, \ and\ \bibinfo {author}
  {\bibfnamefont {D.}~\bibnamefont {Mihailovic}},\ }\href {\doibase
  10.1103/PhysRevLett.101.227001} {\bibfield  {journal} {\bibinfo  {journal}
  {Physical Review Letters}\ }\textbf {\bibinfo {volume} {101}},\ \bibinfo
  {pages} {227001} (\bibinfo {year} {2008})}\BibitemShut {NoStop}%
\bibitem [{\citenamefont {Giannetti}\ \emph {et~al.}(2009)\citenamefont
  {Giannetti}, \citenamefont {Coslovich}, \citenamefont {Cilento},
  \citenamefont {Ferrini}, \citenamefont {Eisaki}, \citenamefont {Kaneko},
  \citenamefont {Greven},\ and\ \citenamefont
  {Parmigiani}}]{giannetti2009discontinuity}%
  \BibitemOpen
  \bibfield  {author} {\bibinfo {author} {\bibfnamefont {C.}~\bibnamefont
  {Giannetti}}, \bibinfo {author} {\bibfnamefont {G.}~\bibnamefont
  {Coslovich}}, \bibinfo {author} {\bibfnamefont {F.}~\bibnamefont {Cilento}},
  \bibinfo {author} {\bibfnamefont {G.}~\bibnamefont {Ferrini}}, \bibinfo
  {author} {\bibfnamefont {H.}~\bibnamefont {Eisaki}}, \bibinfo {author}
  {\bibfnamefont {N.}~\bibnamefont {Kaneko}}, \bibinfo {author} {\bibfnamefont
  {M.}~\bibnamefont {Greven}}, \ and\ \bibinfo {author} {\bibfnamefont
  {F.}~\bibnamefont {Parmigiani}},\ }\href@noop {} {\bibfield  {journal}
  {\bibinfo  {journal} {Physical Review B}\ }\textbf {\bibinfo {volume} {79}},\
  \bibinfo {pages} {224502} (\bibinfo {year} {2009})}\BibitemShut {NoStop}%
\bibitem [{\citenamefont {Toda}\ \emph {et~al.}(2011)\citenamefont {Toda},
  \citenamefont {Mertelj}, \citenamefont {Kusar}, \citenamefont {Kurosawa},
  \citenamefont {Oda}, \citenamefont {Ido},\ and\ \citenamefont
  {Mihailovic}}]{toda2011quasiparticle}%
  \BibitemOpen
  \bibfield  {author} {\bibinfo {author} {\bibfnamefont {Y.}~\bibnamefont
  {Toda}}, \bibinfo {author} {\bibfnamefont {T.}~\bibnamefont {Mertelj}},
  \bibinfo {author} {\bibfnamefont {P.}~\bibnamefont {Kusar}}, \bibinfo
  {author} {\bibfnamefont {T.}~\bibnamefont {Kurosawa}}, \bibinfo {author}
  {\bibfnamefont {M.}~\bibnamefont {Oda}}, \bibinfo {author} {\bibfnamefont
  {M.}~\bibnamefont {Ido}}, \ and\ \bibinfo {author} {\bibfnamefont
  {D.}~\bibnamefont {Mihailovic}},\ }\href@noop {} {\bibfield  {journal}
  {\bibinfo  {journal} {Physical Review B}\ }\textbf {\bibinfo {volume} {84}},\
  \bibinfo {pages} {174516} (\bibinfo {year} {2011})}\BibitemShut {NoStop}%
\bibitem [{\citenamefont {Beyer}\ \emph {et~al.}(2011)\citenamefont {Beyer},
  \citenamefont {St\"adter}, \citenamefont {Beck}, \citenamefont {Sch\"afer},
  \citenamefont {Kabanov}, \citenamefont {Logvenov}, \citenamefont {Bozovic},
  \citenamefont {Koren},\ and\ \citenamefont {Demsar}}]{BeyerStaedter2011}%
  \BibitemOpen
  \bibfield  {author} {\bibinfo {author} {\bibfnamefont {M.}~\bibnamefont
  {Beyer}}, \bibinfo {author} {\bibfnamefont {D.}~\bibnamefont {St\"adter}},
  \bibinfo {author} {\bibfnamefont {M.}~\bibnamefont {Beck}}, \bibinfo {author}
  {\bibfnamefont {H.}~\bibnamefont {Sch\"afer}}, \bibinfo {author}
  {\bibfnamefont {V.~V.}\ \bibnamefont {Kabanov}}, \bibinfo {author}
  {\bibfnamefont {G.}~\bibnamefont {Logvenov}}, \bibinfo {author}
  {\bibfnamefont {I.}~\bibnamefont {Bozovic}}, \bibinfo {author} {\bibfnamefont
  {G.}~\bibnamefont {Koren}}, \ and\ \bibinfo {author} {\bibfnamefont
  {J.}~\bibnamefont {Demsar}},\ }\href {\doibase 10.1103/PhysRevB.83.214515}
  {\bibfield  {journal} {\bibinfo  {journal} {Phys. Rev. B}\ }\textbf {\bibinfo
  {volume} {83}},\ \bibinfo {pages} {214515} (\bibinfo {year}
  {2011})}\BibitemShut {NoStop}%
\bibitem [{\citenamefont {Smallwood}\ \emph {et~al.}(2012)\citenamefont
  {Smallwood}, \citenamefont {Hinton}, \citenamefont {Jozwiak}, \citenamefont
  {Zhang}, \citenamefont {Koralek}, \citenamefont {Eisaki}, \citenamefont
  {Lee}, \citenamefont {Orenstein},\ and\ \citenamefont
  {Lanzara}}]{SmallwoodHinton2012}%
  \BibitemOpen
  \bibfield  {author} {\bibinfo {author} {\bibfnamefont {C.~L.}\ \bibnamefont
  {Smallwood}}, \bibinfo {author} {\bibfnamefont {J.~P.}\ \bibnamefont
  {Hinton}}, \bibinfo {author} {\bibfnamefont {C.}~\bibnamefont {Jozwiak}},
  \bibinfo {author} {\bibfnamefont {W.}~\bibnamefont {Zhang}}, \bibinfo
  {author} {\bibfnamefont {J.~D.}\ \bibnamefont {Koralek}}, \bibinfo {author}
  {\bibfnamefont {H.}~\bibnamefont {Eisaki}}, \bibinfo {author} {\bibfnamefont
  {D.-H.}\ \bibnamefont {Lee}}, \bibinfo {author} {\bibfnamefont
  {J.}~\bibnamefont {Orenstein}}, \ and\ \bibinfo {author} {\bibfnamefont
  {A.}~\bibnamefont {Lanzara}},\ }\href {\doibase 10.1126/science.1217423}
  {\bibfield  {journal} {\bibinfo  {journal} {Science}\ }\textbf {\bibinfo
  {volume} {336}},\ \bibinfo {pages} {1137} (\bibinfo {year} {2012})},\ \Eprint
  {http://arxiv.org/abs/http://science.sciencemag.org/content/336/6085/1137.full.pdf}
  {http://science.sciencemag.org/content/336/6085/1137.full.pdf} \BibitemShut
  {NoStop}%
\bibitem [{\citenamefont {Smallwood}\ \emph {et~al.}(2014)\citenamefont
  {Smallwood}, \citenamefont {Zhang}, \citenamefont {Miller}, \citenamefont
  {Jozwiak}, \citenamefont {Eisaki}, \citenamefont {Lee},\ and\ \citenamefont
  {Lanzara}}]{SmallwoodZhang2014}%
  \BibitemOpen
  \bibfield  {author} {\bibinfo {author} {\bibfnamefont {C.~L.}\ \bibnamefont
  {Smallwood}}, \bibinfo {author} {\bibfnamefont {W.}~\bibnamefont {Zhang}},
  \bibinfo {author} {\bibfnamefont {T.~L.}\ \bibnamefont {Miller}}, \bibinfo
  {author} {\bibfnamefont {C.}~\bibnamefont {Jozwiak}}, \bibinfo {author}
  {\bibfnamefont {H.}~\bibnamefont {Eisaki}}, \bibinfo {author} {\bibfnamefont
  {D.-H.}\ \bibnamefont {Lee}}, \ and\ \bibinfo {author} {\bibfnamefont
  {A.}~\bibnamefont {Lanzara}},\ }\href {\doibase 10.1103/PhysRevB.89.115126}
  {\bibfield  {journal} {\bibinfo  {journal} {Phys. Rev. B}\ }\textbf {\bibinfo
  {volume} {89}},\ \bibinfo {pages} {115126} (\bibinfo {year}
  {2014})}\BibitemShut {NoStop}%
\bibitem [{\citenamefont {Smallwood}\ \emph {et~al.}(2015)\citenamefont
  {Smallwood}, \citenamefont {Zhang}, \citenamefont {Miller}, \citenamefont
  {Affeldt}, \citenamefont {Kurashima}, \citenamefont {Jozwiak}, \citenamefont
  {Noji}, \citenamefont {Koike}, \citenamefont {Eisaki}, \citenamefont {Lee},
  \citenamefont {Kaindl},\ and\ \citenamefont {Lanzara}}]{SamallwoodZhang2015}%
  \BibitemOpen
  \bibfield  {author} {\bibinfo {author} {\bibfnamefont {C.~L.}\ \bibnamefont
  {Smallwood}}, \bibinfo {author} {\bibfnamefont {W.}~\bibnamefont {Zhang}},
  \bibinfo {author} {\bibfnamefont {T.~L.}\ \bibnamefont {Miller}}, \bibinfo
  {author} {\bibfnamefont {G.}~\bibnamefont {Affeldt}}, \bibinfo {author}
  {\bibfnamefont {K.}~\bibnamefont {Kurashima}}, \bibinfo {author}
  {\bibfnamefont {C.}~\bibnamefont {Jozwiak}}, \bibinfo {author} {\bibfnamefont
  {T.}~\bibnamefont {Noji}}, \bibinfo {author} {\bibfnamefont {Y.}~\bibnamefont
  {Koike}}, \bibinfo {author} {\bibfnamefont {H.}~\bibnamefont {Eisaki}},
  \bibinfo {author} {\bibfnamefont {D.-H.}\ \bibnamefont {Lee}}, \bibinfo
  {author} {\bibfnamefont {R.~A.}\ \bibnamefont {Kaindl}}, \ and\ \bibinfo
  {author} {\bibfnamefont {A.}~\bibnamefont {Lanzara}},\ }\href {\doibase
  10.1103/PhysRevB.92.161102} {\bibfield  {journal} {\bibinfo  {journal} {Phys.
  Rev. B}\ }\textbf {\bibinfo {volume} {92}},\ \bibinfo {pages} {161102}
  (\bibinfo {year} {2015})}\BibitemShut {NoStop}%
\bibitem [{\citenamefont {Piovera}\ \emph {et~al.}(2015)\citenamefont
  {Piovera}, \citenamefont {Zhang}, \citenamefont {d'Astuto}, \citenamefont
  {Taleb-Ibrahimi}, \citenamefont {Papalazarou}, \citenamefont {Marsi},
  \citenamefont {Li}, \citenamefont {Raffy},\ and\ \citenamefont
  {Perfetti}}]{PioveraZhang2015}%
  \BibitemOpen
  \bibfield  {author} {\bibinfo {author} {\bibfnamefont {C.}~\bibnamefont
  {Piovera}}, \bibinfo {author} {\bibfnamefont {Z.}~\bibnamefont {Zhang}},
  \bibinfo {author} {\bibfnamefont {M.}~\bibnamefont {d'Astuto}}, \bibinfo
  {author} {\bibfnamefont {A.}~\bibnamefont {Taleb-Ibrahimi}}, \bibinfo
  {author} {\bibfnamefont {E.}~\bibnamefont {Papalazarou}}, \bibinfo {author}
  {\bibfnamefont {M.}~\bibnamefont {Marsi}}, \bibinfo {author} {\bibfnamefont
  {Z.~Z.}\ \bibnamefont {Li}}, \bibinfo {author} {\bibfnamefont
  {H.}~\bibnamefont {Raffy}}, \ and\ \bibinfo {author} {\bibfnamefont
  {L.}~\bibnamefont {Perfetti}},\ }\href {\doibase 10.1103/PhysRevB.91.224509}
  {\bibfield  {journal} {\bibinfo  {journal} {Phys. Rev. B}\ }\textbf {\bibinfo
  {volume} {91}},\ \bibinfo {pages} {224509} (\bibinfo {year}
  {2015})}\BibitemShut {NoStop}%
\bibitem [{Note2()}]{Note2}%
  \BibitemOpen
  \bibinfo {note} {Limited to the vicinity to the nodal point on the $\Gamma
  $-Y line.}\BibitemShut {Stop}%
\bibitem [{\citenamefont {Devereaux}\ and\ \citenamefont
  {Hackl}(2007)}]{DevereauxHackl07}%
  \BibitemOpen
  \bibfield  {author} {\bibinfo {author} {\bibfnamefont {T.~P.}\ \bibnamefont
  {Devereaux}}\ and\ \bibinfo {author} {\bibfnamefont {R.}~\bibnamefont
  {Hackl}},\ }\href@noop {} {\bibfield  {journal} {\bibinfo  {journal} {Reviews
  of Modern Physics}\ }\textbf {\bibinfo {volume} {79}},\ \bibinfo {pages}
  {175} (\bibinfo {year} {2007})}\BibitemShut {NoStop}%
\bibitem [{\citenamefont {Toda}\ \emph {et~al.}(2014)\citenamefont {Toda},
  \citenamefont {Kawanokami}, \citenamefont {Kurosawa}, \citenamefont {Oda},
  \citenamefont {Madan}, \citenamefont {Mertelj}, \citenamefont {Kabanov},\
  and\ \citenamefont {Mihailovic}}]{Toda2014}%
  \BibitemOpen
  \bibfield  {author} {\bibinfo {author} {\bibfnamefont {Y.}~\bibnamefont
  {Toda}}, \bibinfo {author} {\bibfnamefont {F.}~\bibnamefont {Kawanokami}},
  \bibinfo {author} {\bibfnamefont {T.}~\bibnamefont {Kurosawa}}, \bibinfo
  {author} {\bibfnamefont {M.}~\bibnamefont {Oda}}, \bibinfo {author}
  {\bibfnamefont {I.}~\bibnamefont {Madan}}, \bibinfo {author} {\bibfnamefont
  {T.}~\bibnamefont {Mertelj}}, \bibinfo {author} {\bibfnamefont {V.~V.}\
  \bibnamefont {Kabanov}}, \ and\ \bibinfo {author} {\bibfnamefont
  {D.}~\bibnamefont {Mihailovic}},\ }\href {\doibase
  10.1103/PhysRevB.90.094513} {\bibfield  {journal} {\bibinfo  {journal}
  {Physical Review B}\ }\textbf {\bibinfo {volume} {90}},\ \bibinfo {pages}
  {094513} (\bibinfo {year} {2014})}\BibitemShut {NoStop}%
\bibitem [{\citenamefont {Coslovich}\ \emph {et~al.}(2013)\citenamefont
  {Coslovich}, \citenamefont {Giannetti}, \citenamefont {Cilento},
  \citenamefont {Dal~Conte}, \citenamefont {Abebaw}, \citenamefont {Bossini},
  \citenamefont {Ferrini}, \citenamefont {Eisaki}, \citenamefont {Greven},
  \citenamefont {Damascelli},\ and\ \citenamefont
  {Parmigiani}}]{CoslovichGianetti2013}%
  \BibitemOpen
  \bibfield  {author} {\bibinfo {author} {\bibfnamefont {G.}~\bibnamefont
  {Coslovich}}, \bibinfo {author} {\bibfnamefont {C.}~\bibnamefont
  {Giannetti}}, \bibinfo {author} {\bibfnamefont {F.}~\bibnamefont {Cilento}},
  \bibinfo {author} {\bibfnamefont {S.}~\bibnamefont {Dal~Conte}}, \bibinfo
  {author} {\bibfnamefont {T.}~\bibnamefont {Abebaw}}, \bibinfo {author}
  {\bibfnamefont {D.}~\bibnamefont {Bossini}}, \bibinfo {author} {\bibfnamefont
  {G.}~\bibnamefont {Ferrini}}, \bibinfo {author} {\bibfnamefont
  {H.}~\bibnamefont {Eisaki}}, \bibinfo {author} {\bibfnamefont
  {M.}~\bibnamefont {Greven}}, \bibinfo {author} {\bibfnamefont
  {A.}~\bibnamefont {Damascelli}}, \ and\ \bibinfo {author} {\bibfnamefont
  {F.}~\bibnamefont {Parmigiani}},\ }\href {\doibase
  10.1103/PhysRevLett.110.107003} {\bibfield  {journal} {\bibinfo  {journal}
  {Phys. Rev. Lett.}\ }\textbf {\bibinfo {volume} {110}},\ \bibinfo {pages}
  {107003} (\bibinfo {year} {2013})}\BibitemShut {NoStop}%
\bibitem [{Note3()}]{Note3}%
  \BibitemOpen
  \bibinfo {note} {Despite Bi2212 is orthorhombic we use the ideal D$_{\protect
  \mathrm {4h}}$ point group tetragonal CuO$_{2}$-plane symmetry to simplify
  the notation. See supplemental to Ref. {[}\protect \rev@citealpnum
  {Toda2014}{]} for details.}\BibitemShut {Stop}%
\bibitem [{\citenamefont {Kabanov}\ \emph {et~al.}(1999)\citenamefont
  {Kabanov}, \citenamefont {Demsar}, \citenamefont {Podobnik},\ and\
  \citenamefont {Mihailovic}}]{KabanovDemsar1999}%
  \BibitemOpen
  \bibfield  {author} {\bibinfo {author} {\bibfnamefont {V.~V.}\ \bibnamefont
  {Kabanov}}, \bibinfo {author} {\bibfnamefont {J.}~\bibnamefont {Demsar}},
  \bibinfo {author} {\bibfnamefont {B.}~\bibnamefont {Podobnik}}, \ and\
  \bibinfo {author} {\bibfnamefont {D.}~\bibnamefont {Mihailovic}},\ }\href
  {\doibase 10.1103/PhysRevB.59.1497} {\bibfield  {journal} {\bibinfo
  {journal} {Phys. Rev. B}\ }\textbf {\bibinfo {volume} {59}},\ \bibinfo
  {pages} {1497} (\bibinfo {year} {1999})}\BibitemShut {NoStop}%
\bibitem [{\citenamefont {Yusupov}\ \emph {et~al.}(2010)\citenamefont
  {Yusupov}, \citenamefont {Mertelj}, \citenamefont {Kabanov}, \citenamefont
  {Brazovskii}, \citenamefont {Kusar}, \citenamefont {Chu}, \citenamefont
  {Fisher},\ and\ \citenamefont {Mihailovic}}]{Yusupov2010}%
  \BibitemOpen
  \bibfield  {author} {\bibinfo {author} {\bibfnamefont {R.}~\bibnamefont
  {Yusupov}}, \bibinfo {author} {\bibfnamefont {T.}~\bibnamefont {Mertelj}},
  \bibinfo {author} {\bibfnamefont {V.~V.}\ \bibnamefont {Kabanov}}, \bibinfo
  {author} {\bibfnamefont {S.}~\bibnamefont {Brazovskii}}, \bibinfo {author}
  {\bibfnamefont {P.}~\bibnamefont {Kusar}}, \bibinfo {author} {\bibfnamefont
  {J.-H.}\ \bibnamefont {Chu}}, \bibinfo {author} {\bibfnamefont {I.~R.}\
  \bibnamefont {Fisher}}, \ and\ \bibinfo {author} {\bibfnamefont
  {D.}~\bibnamefont {Mihailovic}},\ }\href {\doibase 10.1038/nphys1738}
  {\bibfield  {journal} {\bibinfo  {journal} {Nature Physics}\ }\textbf
  {\bibinfo {volume} {6}},\ \bibinfo {pages} {681} (\bibinfo {year}
  {2010})}\BibitemShut {NoStop}%
\bibitem [{\citenamefont {Madan}\ \emph {et~al.}(2016)\citenamefont {Madan},
  \citenamefont {Kusar}, \citenamefont {Baranov}, \citenamefont {Lu-Dac},
  \citenamefont {Kabanov}, \citenamefont {Mertelj},\ and\ \citenamefont
  {Mihailovic}}]{Madan2016}%
  \BibitemOpen
  \bibfield  {author} {\bibinfo {author} {\bibfnamefont {I.}~\bibnamefont
  {Madan}}, \bibinfo {author} {\bibfnamefont {P.}~\bibnamefont {Kusar}},
  \bibinfo {author} {\bibfnamefont {V.~V.}\ \bibnamefont {Baranov}}, \bibinfo
  {author} {\bibfnamefont {M.}~\bibnamefont {Lu-Dac}}, \bibinfo {author}
  {\bibfnamefont {V.~V.}\ \bibnamefont {Kabanov}}, \bibinfo {author}
  {\bibfnamefont {T.}~\bibnamefont {Mertelj}}, \ and\ \bibinfo {author}
  {\bibfnamefont {D.}~\bibnamefont {Mihailovic}},\ }\href {\doibase
  10.1103/PhysRevB.93.224520} {\bibfield  {journal} {\bibinfo  {journal}
  {Physical Review B}\ }\textbf {\bibinfo {volume} {93}},\ \bibinfo {pages}
  {224520} (\bibinfo {year} {2016})}\BibitemShut {NoStop}%
\bibitem [{\citenamefont {Madan}\ \emph {et~al.}(2015)\citenamefont {Madan},
  \citenamefont {Kurosawa}, \citenamefont {Toda}, \citenamefont {Oda},
  \citenamefont {Mertelj},\ and\ \citenamefont {Mihailovic}}]{Madan2015}%
  \BibitemOpen
  \bibfield  {author} {\bibinfo {author} {\bibfnamefont {I.}~\bibnamefont
  {Madan}}, \bibinfo {author} {\bibfnamefont {T.}~\bibnamefont {Kurosawa}},
  \bibinfo {author} {\bibfnamefont {Y.}~\bibnamefont {Toda}}, \bibinfo {author}
  {\bibfnamefont {M.}~\bibnamefont {Oda}}, \bibinfo {author} {\bibfnamefont
  {T.}~\bibnamefont {Mertelj}}, \ and\ \bibinfo {author} {\bibfnamefont
  {D.}~\bibnamefont {Mihailovic}},\ }\href {\doibase 10.1038/ncomms7958}
  {\bibfield  {journal} {\bibinfo  {journal} {Nature Communications}\ }\textbf
  {\bibinfo {volume} {6}},\ \bibinfo {pages} {6958} (\bibinfo {year} {2015})},\
  \Eprint {http://arxiv.org/abs/1410.3205} {arXiv:1410.3205} \BibitemShut
  {NoStop}%
\bibitem [{\citenamefont {Cort\'es}\ \emph {et~al.}(2011)\citenamefont
  {Cort\'es}, \citenamefont {Rettig}, \citenamefont {Yoshida}, \citenamefont
  {Eisaki}, \citenamefont {Wolf},\ and\ \citenamefont
  {Bovensiepen}}]{CortesRettig2011}%
  \BibitemOpen
  \bibfield  {author} {\bibinfo {author} {\bibfnamefont {R.}~\bibnamefont
  {Cort\'es}}, \bibinfo {author} {\bibfnamefont {L.}~\bibnamefont {Rettig}},
  \bibinfo {author} {\bibfnamefont {Y.}~\bibnamefont {Yoshida}}, \bibinfo
  {author} {\bibfnamefont {H.}~\bibnamefont {Eisaki}}, \bibinfo {author}
  {\bibfnamefont {M.}~\bibnamefont {Wolf}}, \ and\ \bibinfo {author}
  {\bibfnamefont {U.}~\bibnamefont {Bovensiepen}},\ }\href {\doibase
  10.1103/PhysRevLett.107.097002} {\bibfield  {journal} {\bibinfo  {journal}
  {Phys. Rev. Lett.}\ }\textbf {\bibinfo {volume} {107}},\ \bibinfo {pages}
  {097002} (\bibinfo {year} {2011})}\BibitemShut {NoStop}%
\bibitem [{\citenamefont {Stojchevska}\ \emph {et~al.}(2011)\citenamefont
  {Stojchevska}, \citenamefont {Kusar}, \citenamefont {Mertelj}, \citenamefont
  {Kabanov}, \citenamefont {Toda}, \citenamefont {Yao},\ and\ \citenamefont
  {Mihailovic}}]{stojchevska2011mechanisms}%
  \BibitemOpen
  \bibfield  {author} {\bibinfo {author} {\bibfnamefont {L.}~\bibnamefont
  {Stojchevska}}, \bibinfo {author} {\bibfnamefont {P.}~\bibnamefont {Kusar}},
  \bibinfo {author} {\bibfnamefont {T.}~\bibnamefont {Mertelj}}, \bibinfo
  {author} {\bibfnamefont {V.}~\bibnamefont {Kabanov}}, \bibinfo {author}
  {\bibfnamefont {Y.}~\bibnamefont {Toda}}, \bibinfo {author} {\bibfnamefont
  {X.}~\bibnamefont {Yao}}, \ and\ \bibinfo {author} {\bibfnamefont
  {D.}~\bibnamefont {Mihailovic}},\ }\href@noop {} {\bibfield  {journal}
  {\bibinfo  {journal} {Physical Review B}\ }\textbf {\bibinfo {volume} {84}},\
  \bibinfo {pages} {180507} (\bibinfo {year} {2011})}\BibitemShut {NoStop}%
\bibitem [{Note4()}]{Note4}%
  \BibitemOpen
  \bibinfo {note} {As in Ref. \protect \rev@citealpnum {Toda2014} we use the
  approximate notation corresponding to the tetragonal symmetry.}\BibitemShut
  {Stop}%
\bibitem [{Note5()}]{Note5}%
  \BibitemOpen
  \bibinfo {note} {The scattering from the D-beam does not contribute
  significantly since the beam is not modulated.}\BibitemShut {Stop}%
\bibitem [{\citenamefont {Perfetti}\ \emph {et~al.}(2007)\citenamefont
  {Perfetti}, \citenamefont {Loukakos}, \citenamefont {Lisowski}, \citenamefont
  {Bovensiepen}, \citenamefont {Eisaki},\ and\ \citenamefont
  {Wolf}}]{Perfetti2007}%
  \BibitemOpen
  \bibfield  {author} {\bibinfo {author} {\bibfnamefont {L.}~\bibnamefont
  {Perfetti}}, \bibinfo {author} {\bibfnamefont {P.~A.}\ \bibnamefont
  {Loukakos}}, \bibinfo {author} {\bibfnamefont {M.}~\bibnamefont {Lisowski}},
  \bibinfo {author} {\bibfnamefont {U.}~\bibnamefont {Bovensiepen}}, \bibinfo
  {author} {\bibfnamefont {H.}~\bibnamefont {Eisaki}}, \ and\ \bibinfo {author}
  {\bibfnamefont {M.}~\bibnamefont {Wolf}},\ }\href {\doibase
  10.1103/PhysRevLett.99.197001} {\bibfield  {journal} {\bibinfo  {journal}
  {Phys. Rev. Lett.}\ }\textbf {\bibinfo {volume} {99}},\ \bibinfo {pages}
  {197001} (\bibinfo {year} {2007})}\BibitemShut {NoStop}%
\bibitem [{Note6()}]{Note6}%
  \BibitemOpen
  \bibinfo {note} {In the vicinity of the peak of the unperturbed $\Delta
  R_{\protect \mathrm {B_{1g}}}/R$-response}\BibitemShut {NoStop}%
\bibitem [{\citenamefont {Kabanov}\ \emph {et~al.}(2005)\citenamefont
  {Kabanov}, \citenamefont {Demsar},\ and\ \citenamefont
  {Mihailovic}}]{Kabanov2005}%
  \BibitemOpen
  \bibfield  {author} {\bibinfo {author} {\bibfnamefont {V.~V.}\ \bibnamefont
  {Kabanov}}, \bibinfo {author} {\bibfnamefont {J.}~\bibnamefont {Demsar}}, \
  and\ \bibinfo {author} {\bibfnamefont {D.}~\bibnamefont {Mihailovic}},\
  }\href {\doibase 10.1103/PhysRevLett.95.147002} {\bibfield  {journal}
  {\bibinfo  {journal} {Physical Review Letters}\ }\textbf {\bibinfo {volume}
  {95}},\ \bibinfo {pages} {147002} (\bibinfo {year} {2005})}\BibitemShut
  {NoStop}%
\bibitem [{\citenamefont {Ku{\v{s}}ar}\ \emph {et~al.}(2011)\citenamefont
  {Ku{\v{s}}ar}, \citenamefont {Kabanov}, \citenamefont {Sugai}, \citenamefont
  {Dem{\v{s}}ar}, \citenamefont {Mertelj},\ and\ \citenamefont
  {Mihailovi{\'c}}}]{kuvsar2011dynamical}%
  \BibitemOpen
  \bibfield  {author} {\bibinfo {author} {\bibfnamefont {P.}~\bibnamefont
  {Ku{\v{s}}ar}}, \bibinfo {author} {\bibfnamefont {V.~V.}\ \bibnamefont
  {Kabanov}}, \bibinfo {author} {\bibfnamefont {S.}~\bibnamefont {Sugai}},
  \bibinfo {author} {\bibfnamefont {J.}~\bibnamefont {Dem{\v{s}}ar}}, \bibinfo
  {author} {\bibfnamefont {T.}~\bibnamefont {Mertelj}}, \ and\ \bibinfo
  {author} {\bibfnamefont {D.}~\bibnamefont {Mihailovi{\'c}}},\ }\href@noop {}
  {\bibfield  {journal} {\bibinfo  {journal} {Journal of superconductivity and
  novel magnetism}\ }\textbf {\bibinfo {volume} {24}},\ \bibinfo {pages} {421}
  (\bibinfo {year} {2011})}\BibitemShut {NoStop}%
\bibitem [{Note7()}]{Note7}%
  \BibitemOpen
  \bibinfo {note} {The $\protect \mathrm {A}_{1g}+\protect \mathrm {B}_{1g}$
  $\Delta R/R$ has a peak at a slightly earlier time, where the PG component
  contribution is significantly larger.}\BibitemShut {Stop}%
\bibitem [{Note8()}]{Note8}%
  \BibitemOpen
  \bibinfo {note} {The difference in sampling time of 0.55 ps needs to be taken
  into account when directly comparing the traces.}\BibitemShut {Stop}%
\bibitem [{Note9()}]{Note9}%
  \BibitemOpen
  \bibinfo {note} {The fully recovered PG component contributes to a
  $t_{\protect \mathrm {D-P}}$-independent negative shift at longer
  $t_{\protect \mathrm {D-P}}$.}\BibitemShut {Stop}%
\bibitem [{Note10()}]{Note10}%
  \BibitemOpen
  \bibinfo {note} {When the negative PG component is suppressed the total
  $\Delta R/R$ increases.}\BibitemShut {Stop}%
\bibitem [{\citenamefont {Kusar}(2007)}]{kusthesis}%
  \BibitemOpen
  \bibfield  {author} {\bibinfo {author} {\bibfnamefont {P.}~\bibnamefont
  {Kusar}},\ }\emph {\bibinfo {title} {Influence of Irregularities and
  Dimensionality on Electron Relaxation}},\ \href@noop {} {Ph.D. thesis},\
  \bibinfo  {school} {Faculty of Mathemathics and Physics, University of
  Ljubljana, Slovenia} (\bibinfo {year} {2007})\BibitemShut {NoStop}%
\bibitem [{Note11()}]{Note11}%
  \BibitemOpen
  \bibinfo {note} {Despite the worse fit quality.}\BibitemShut {Stop}%
\bibitem [{Note12()}]{Note12}%
  \BibitemOpen
  \bibinfo {note} {An consistent increase of the QP relaxation time away from
  the nodal point was observed in a recent ARPES experiment\cite
  {SamallwoodZhang2015} in some controversy to an earlier ARPES result\cite
  {CortesRettig2011}.}\BibitemShut {Stop}%
\bibitem [{\citenamefont {Smallwood}\ \emph {et~al.}(2016)\citenamefont
  {Smallwood}, \citenamefont {Miller}, \citenamefont {Zhang}, \citenamefont
  {Kaindl},\ and\ \citenamefont {Lanzara}}]{SmallwoodMiller2016}%
  \BibitemOpen
  \bibfield  {author} {\bibinfo {author} {\bibfnamefont {C.~L.}\ \bibnamefont
  {Smallwood}}, \bibinfo {author} {\bibfnamefont {T.~L.}\ \bibnamefont
  {Miller}}, \bibinfo {author} {\bibfnamefont {W.}~\bibnamefont {Zhang}},
  \bibinfo {author} {\bibfnamefont {R.~A.}\ \bibnamefont {Kaindl}}, \ and\
  \bibinfo {author} {\bibfnamefont {A.}~\bibnamefont {Lanzara}},\ }\href
  {\doibase 10.1103/PhysRevB.93.235107} {\bibfield  {journal} {\bibinfo
  {journal} {Phys. Rev. B}\ }\textbf {\bibinfo {volume} {93}},\ \bibinfo
  {pages} {235107} (\bibinfo {year} {2016})}\BibitemShut {NoStop}%
\bibitem [{Note13()}]{Note13}%
  \BibitemOpen
  \bibinfo {note} {In LSCO the lattice temperature after the destruction pulse
  reaches\cite {Madan2016} $T_{\protect \mathrm {c}}$ at $\protect \mathcal
  {F}_{\protect \mathrm {D}}=20$$\mu $J/cm$^{2}$.}\BibitemShut {Stop}%
\bibitem [{\citenamefont {Zhang}\ \emph {et~al.}(2013)\citenamefont {Zhang},
  \citenamefont {Smallwood}, \citenamefont {Jozwiak}, \citenamefont {Miller},
  \citenamefont {Yoshida}, \citenamefont {Eisaki}, \citenamefont {Lee},\ and\
  \citenamefont {Lanzara}}]{ZhangSmallwood2013}%
  \BibitemOpen
  \bibfield  {author} {\bibinfo {author} {\bibfnamefont {W.}~\bibnamefont
  {Zhang}}, \bibinfo {author} {\bibfnamefont {C.~L.}\ \bibnamefont
  {Smallwood}}, \bibinfo {author} {\bibfnamefont {C.}~\bibnamefont {Jozwiak}},
  \bibinfo {author} {\bibfnamefont {T.~L.}\ \bibnamefont {Miller}}, \bibinfo
  {author} {\bibfnamefont {Y.}~\bibnamefont {Yoshida}}, \bibinfo {author}
  {\bibfnamefont {H.}~\bibnamefont {Eisaki}}, \bibinfo {author} {\bibfnamefont
  {D.-H.}\ \bibnamefont {Lee}}, \ and\ \bibinfo {author} {\bibfnamefont
  {A.}~\bibnamefont {Lanzara}},\ }\href {\doibase 10.1103/PhysRevB.88.245132}
  {\bibfield  {journal} {\bibinfo  {journal} {Phys. Rev. B}\ }\textbf {\bibinfo
  {volume} {88}},\ \bibinfo {pages} {245132} (\bibinfo {year}
  {2013})}\BibitemShut {NoStop}%
\bibitem [{\citenamefont {Madan}\ \emph {et~al.}(2014)\citenamefont {Madan},
  \citenamefont {Kurosawa}, \citenamefont {Toda}, \citenamefont {Oda},
  \citenamefont {Mertelj}, \citenamefont {Kusar},\ and\ \citenamefont
  {Mihailovic}}]{madan2014separating}%
  \BibitemOpen
  \bibfield  {author} {\bibinfo {author} {\bibfnamefont {I.}~\bibnamefont
  {Madan}}, \bibinfo {author} {\bibfnamefont {T.}~\bibnamefont {Kurosawa}},
  \bibinfo {author} {\bibfnamefont {Y.}~\bibnamefont {Toda}}, \bibinfo {author}
  {\bibfnamefont {M.}~\bibnamefont {Oda}}, \bibinfo {author} {\bibfnamefont
  {T.}~\bibnamefont {Mertelj}}, \bibinfo {author} {\bibfnamefont
  {P.}~\bibnamefont {Kusar}}, \ and\ \bibinfo {author} {\bibfnamefont
  {D.}~\bibnamefont {Mihailovic}},\ }\href@noop {} {\bibfield  {journal}
  {\bibinfo  {journal} {Scientific Reports}\ }\textbf {\bibinfo {volume} {4}},\
  \bibinfo {pages} {5656} (\bibinfo {year} {2014})}\BibitemShut {NoStop}%
\bibitem [{Note14()}]{Note14}%
  \BibitemOpen
  \bibinfo {note} {The optical penetration depth in Bi2212\cite
  {stojchevska2011mechanisms} is of the order of 100 nm in comparison to a nm
  scale photelectron escape depth.}\BibitemShut {Stop}%
\bibitem [{Note15()}]{Note15}%
  \BibitemOpen
  \bibinfo {note} {The B$_{1\protect \mathrm {g}}$ optical response is
  sensitive to a broad region near the anti node while the A$_{1\protect
  \mathrm {g}}$ response samples both the nodal and antinodal regions\cite
  {DevereauxHackl07}}\BibitemShut {NoStop}%
\bibitem [{Note16()}]{Note16}%
  \BibitemOpen
  \bibinfo {note} {Due to the exponential decay of the excitation fluence away
  from the surface the equivalent external in all-optical experiment is $\sim
  1.5$ times larger than in the case of TR-ARPES.}\BibitemShut {Stop}%
\bibitem [{Note17()}]{Note17}%
  \BibitemOpen
  \bibinfo {note} {The surface gap in the case of TR-ARPES and the bulk gap in
  the case of optics.}\BibitemShut {Stop}%
\bibitem [{\citenamefont {Kibble}\ and\ \citenamefont
  {Volovik}(1997)}]{Kibble1997}%
  \BibitemOpen
  \bibfield  {author} {\bibinfo {author} {\bibfnamefont {T.~W.~B.}\
  \bibnamefont {Kibble}}\ and\ \bibinfo {author} {\bibfnamefont {G.~E.}\
  \bibnamefont {Volovik}},\ }\href {\doibase 10.1134/1.567332} {\bibfield
  {journal} {\bibinfo  {journal} {Journal of Experimental and Theoretical
  Physics Letters}\ }\textbf {\bibinfo {volume} {65}},\ \bibinfo {pages} {102}
  (\bibinfo {year} {1997})}\BibitemShut {NoStop}%
\bibitem [{\citenamefont {Zurek}(1996)}]{Zurek1996}%
  \BibitemOpen
  \bibfield  {author} {\bibinfo {author} {\bibfnamefont {W.}~\bibnamefont
  {Zurek}},\ }\href {\doibase 10.1016/S0370-1573(96)00009-9} {\bibfield
  {journal} {\bibinfo  {journal} {Physics Reports}\ }\textbf {\bibinfo {volume}
  {276}},\ \bibinfo {pages} {177} (\bibinfo {year} {1996})}\BibitemShut
  {NoStop}%
\bibitem [{Note18()}]{Note18}%
  \BibitemOpen
  \bibinfo {note} {Supplemental information to Ref. {[}\protect \rev@citealpnum
  {Madan2016}{]}}\BibitemShut {NoStop}%
\end{thebibliography}%

\end{document}